\documentclass[article,12pt]{elsarticle}
\biboptions{sort&compress}
\usepackage{graphicx}
\usepackage{subcaption}
\usepackage{float}
\usepackage{amssymb}
\usepackage{amsmath}
\usepackage{booktabs}
\usepackage{verbatim}
\usepackage{adjustbox}
\usepackage{url}

\usepackage{xcolor}
\usepackage{caption}
\usepackage{enumitem}
\usepackage{hyperref}
\hypersetup{
    colorlinks=true,
    linkcolor=blue,
    filecolor=magenta,      
    urlcolor=cyan,
}
\journal{Powder Technology}
\begin{document}
\begin{frontmatter}

\title{Calibration of a DEM contact model for wet industrial granular materials}
\author[msm]{Sahar Pourandi\corref{cor1}}
\ead{s.pourandi@utwente.nl}

\author[dpm]{P. Christian van der Sande}
\ead{c.vandersande@hmbv.hosokawa.com}
\author[msm]{Igor Ostanin}
\ead{i.ostanin@utwente.nl}
\author[msm]{Thomas Weinhart}
\ead{t.weinhart@utwente.nl}

\cortext[cor1]{Corresponding author.}

\address[msm]{University of Twente, Drienerlolaan 5, Enschede, The Netherlands}
\address[dpm]{Delft University of Technology, Van der Maasweg 9, Delft, The Netherlands}
%------------------------------
\begin{abstract}
This study presents and calibrates a Discrete Element Method (DEM) contact model for wet granular materials in the pendular regime. The model extends a previously calibrated dry contact formulation by incorporating liquid bridges that generate capillary adhesion between particles, while liquid migration is represented through evolving bridge volumes. Two reactor-grade polypropylene powders with different particle size distributions, bulk densities, and surface morphologies are investigated, resulting in distinct wetting behavior.
A schematic framework is introduced to relate increasing liquid content to the transition from dry to wet contacts using two key parameters: the minimum liquid film volume and the maximum liquid bridge volume. These parameters are calibrated using dynamic angle of repose measurements from rotating drum experiments. The calibrated model reproduces the experimental flow behavior of both powders: full agreement is obtained for the coarser, more porous powder across all liquid contents, while for the finer, denser powder, agreement is achieved at low to moderate liquid contents. At higher liquid contents, discrepancies arise due to agglomeration effects amplified by particle scaling.
These results demonstrate the effectiveness of the dynamic angle of repose as a calibration target and highlight the limitations of particle scaling for strongly cohesive wet granular systems. The proposed framework provides a practical basis for DEM-based modeling of wet powder flow in industrial processes.
\end{abstract}


\end{frontmatter}

%\linenumbers
%
%\textcolor{red}{Pls. mark changes by one color, e.g., blue - and exchange all apostrophs to proper ones ``text'', change e.g. to e.g.\ TEXT, as well as i.e. to i.e.\ TEXT ... thanks!}
\section{Introduction}
\begin{figure}[H]
	\centering
	\includegraphics[width=\textwidth]{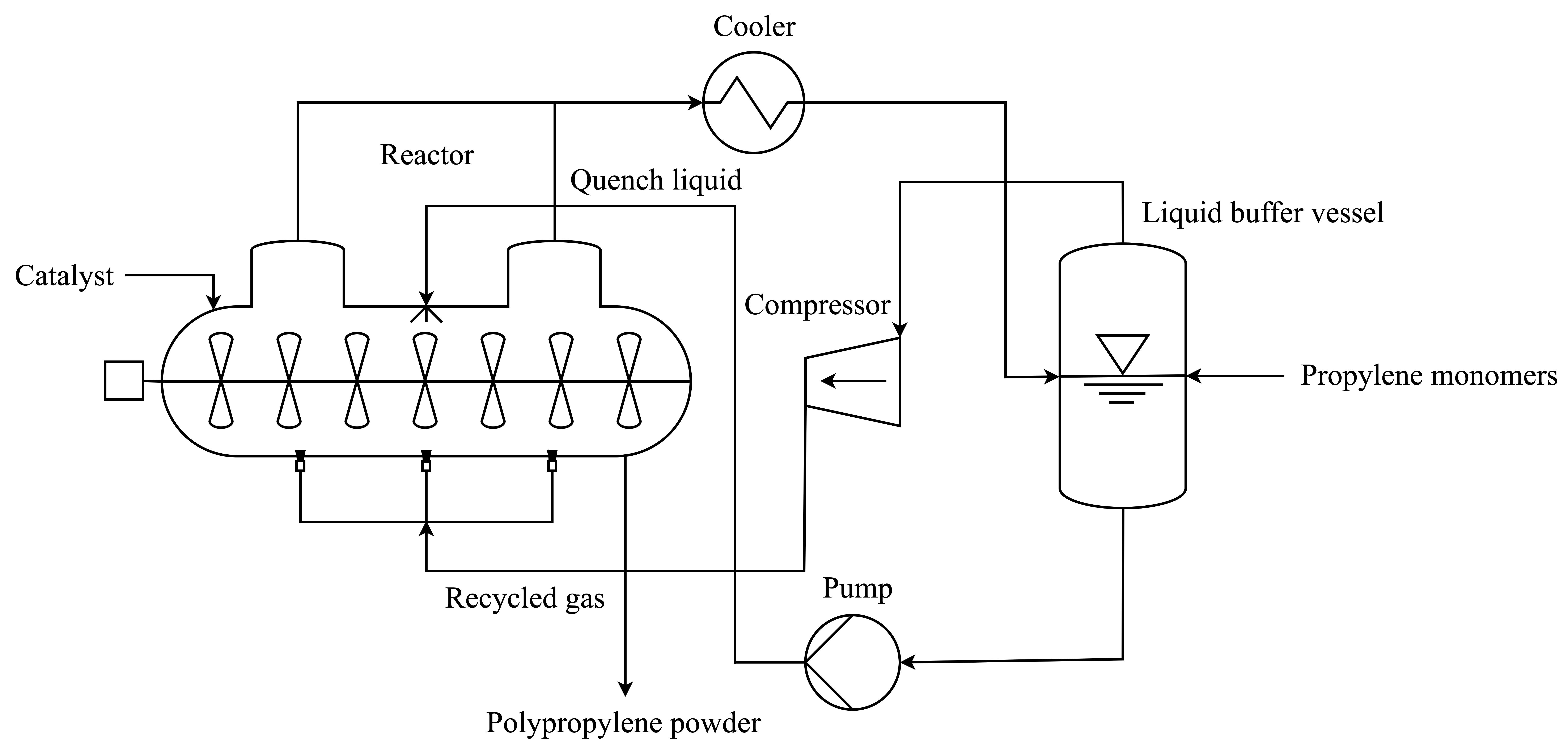}
	\caption{Schematic illustration of the Innovene polypropylene (PP) process, adapted from \cite{dittrich2007residence}.}
	\label{fig:HSBR-Schematic}
\end{figure}

In industrial processes, granular materials are frequently exposed to small amounts of liquid, which can fundamentally alter their flow behavior. In gas-phase polymerization, for example, quench liquids – such as liquid propylene monomer – are sprayed onto the surface of the powder in limited amounts to remove heat generated during polymerization (Figure~\ref{fig:HSBR-Schematic}) \cite{dittrich2007residence, soares2013polyolefin}. Beyond cooling, the added liquid also affects powder flow. In particular, non-uniform liquid distribution can create locally over-wetted regions, which promotes liquid bridging and the formation of agglomerates or lumps. Such lump formation may lead to obstructions in the discharging pipeline, disruptions in heat exchange, and deviations in flow patterns. In severe cases, these issues can lead to unscheduled plant shutdowns \cite{van2024flow}.
Recent experimental studies using X-ray imaging have demonstrated that the presence of liquid significantly influences the dynamics of polypropylene reactor powders in horizontal stirred bed reactors (HSBRs), increasing cohesion and reducing flowability due to liquid bridging at particle contacts \cite{van2024flow}.
These observations highlight the importance of uniform wetting inside the reactor and motivate the inclusion of liquid migration effects in DEM, since the redistribution of liquid between particles directly governs the development of cohesive regions.

\begin{figure}[H]
	\centering
	\renewcommand{\arraystretch}{1} % optional: adds spacing between rows
	\begin{tabular}{c@{\hspace{4pt}}c@{\hspace{4pt}}c@{\hspace{4pt}}c@{\hspace{4pt}}c}
		\raisebox{-.5\height}{\includegraphics[width=0.16\textwidth]{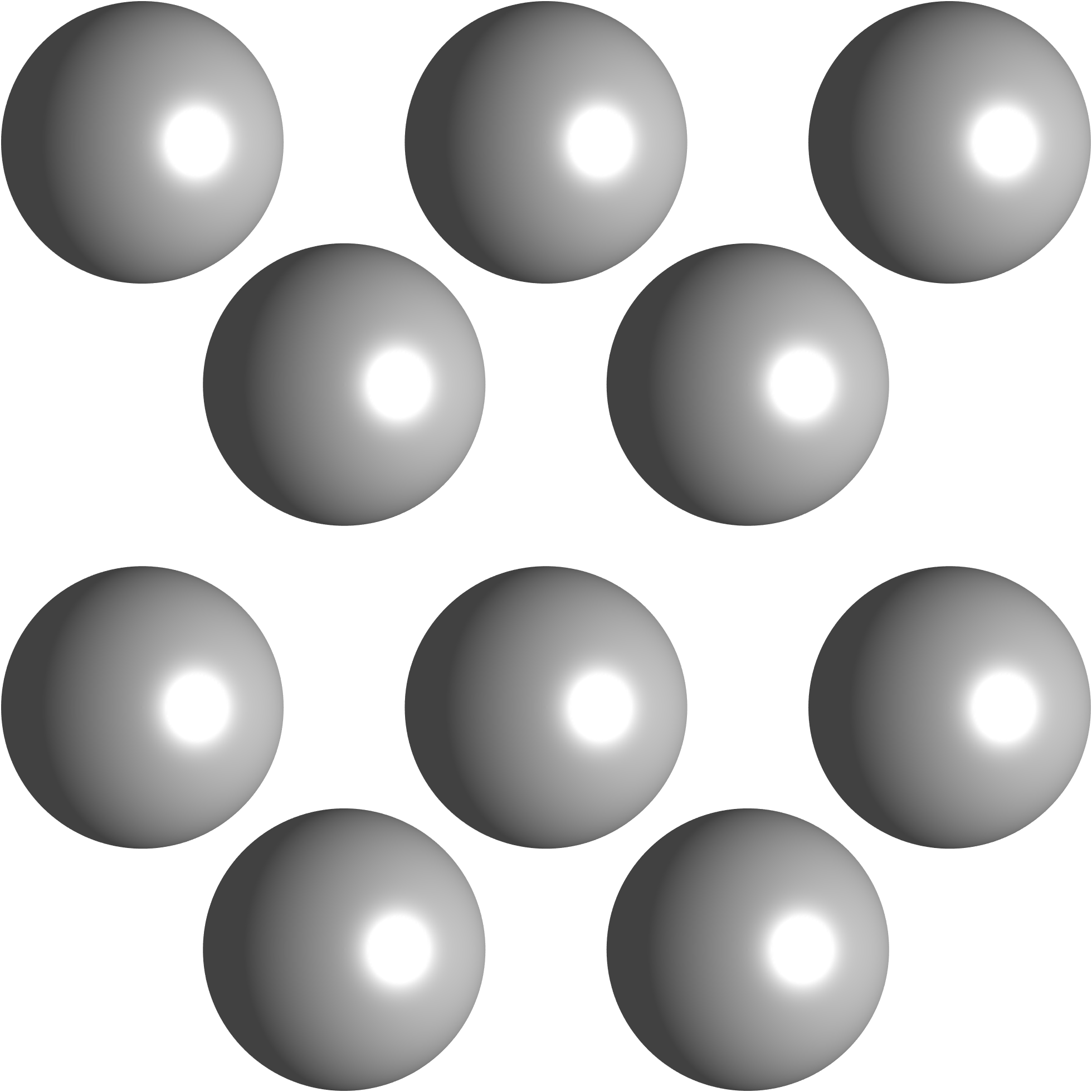}} &
		\raisebox{-.5\height}{\includegraphics[width=0.16\textwidth]{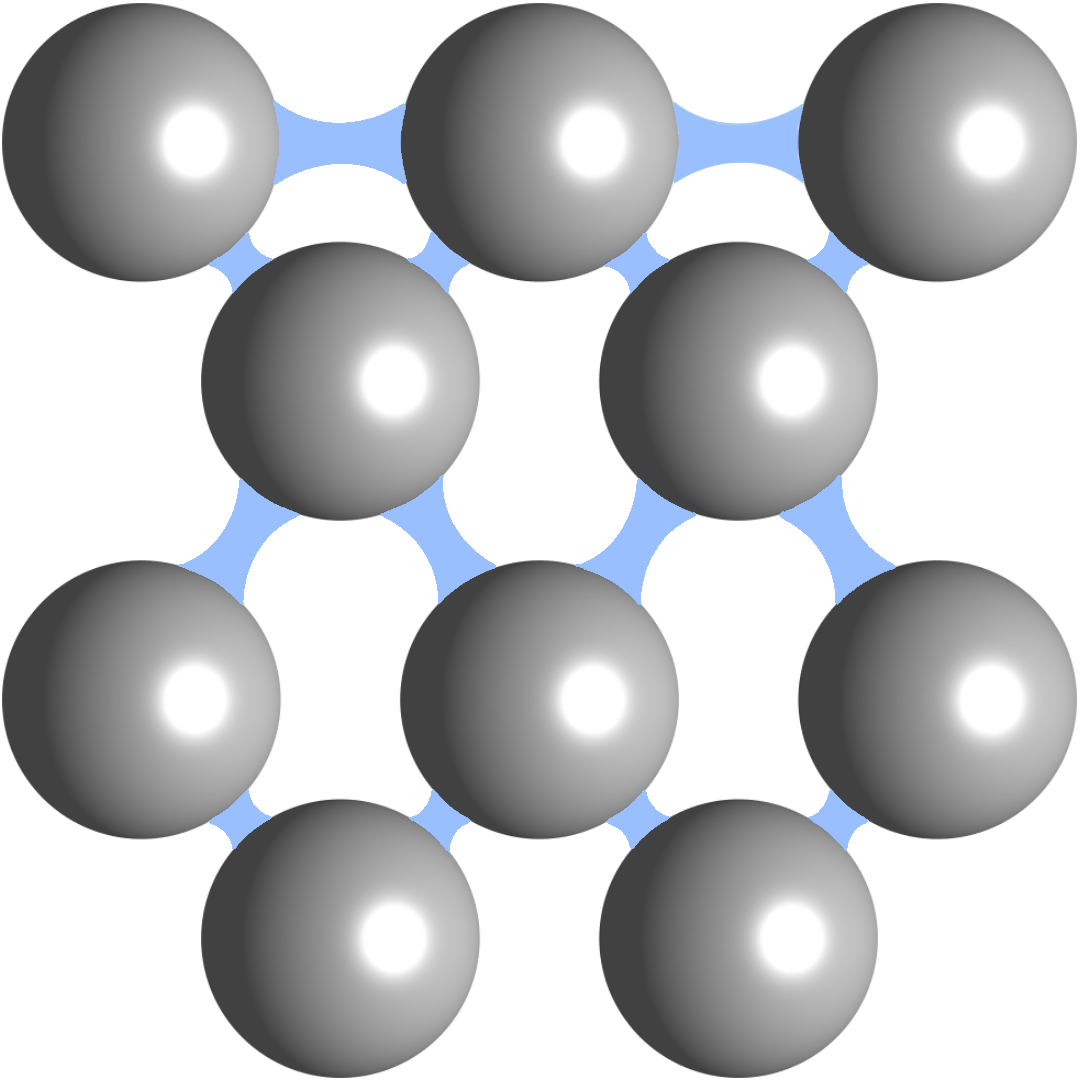}} &
		\raisebox{-.5\height}{\includegraphics[width=0.16\textwidth]{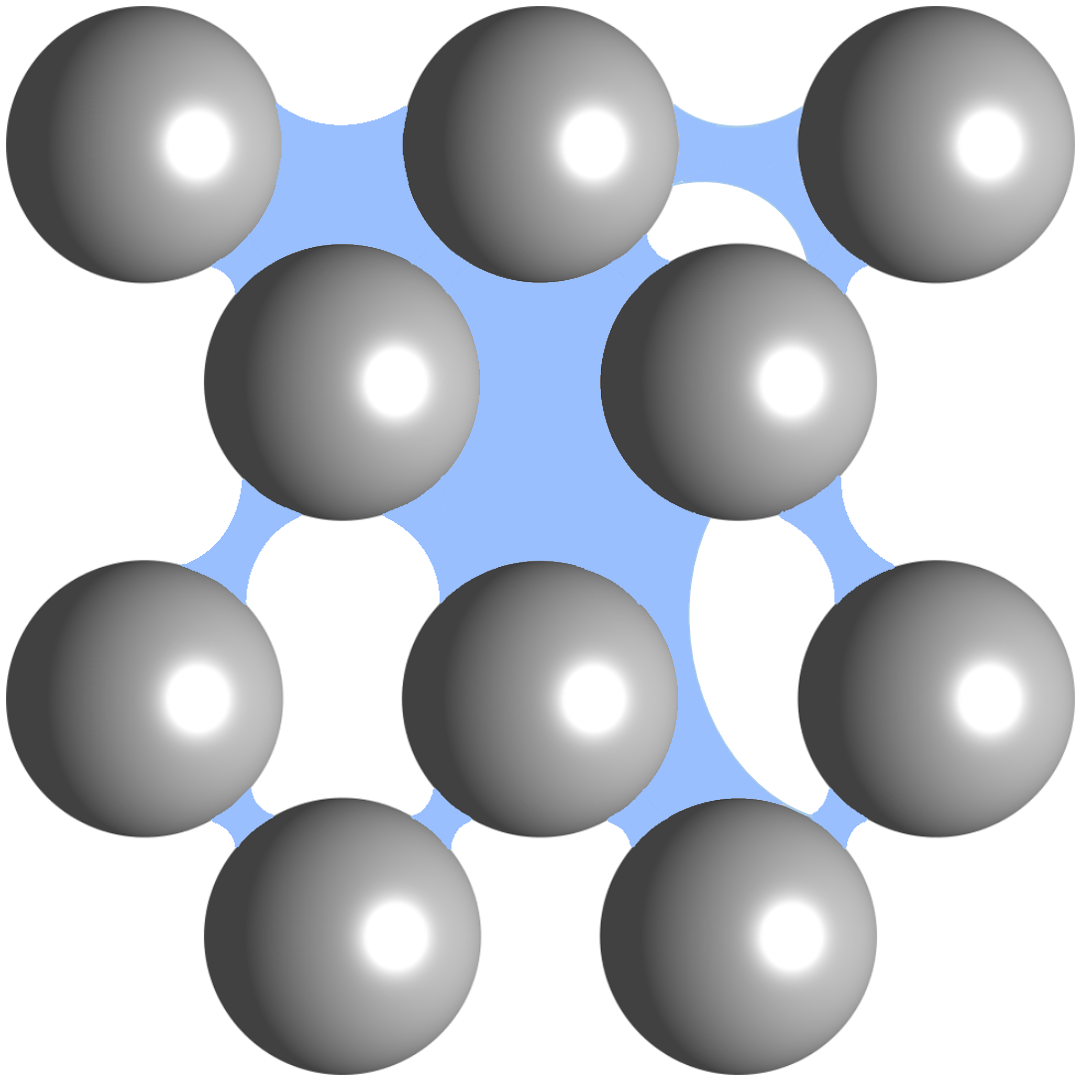}} &
		\raisebox{-.5\height}{\includegraphics[width=0.16\textwidth]{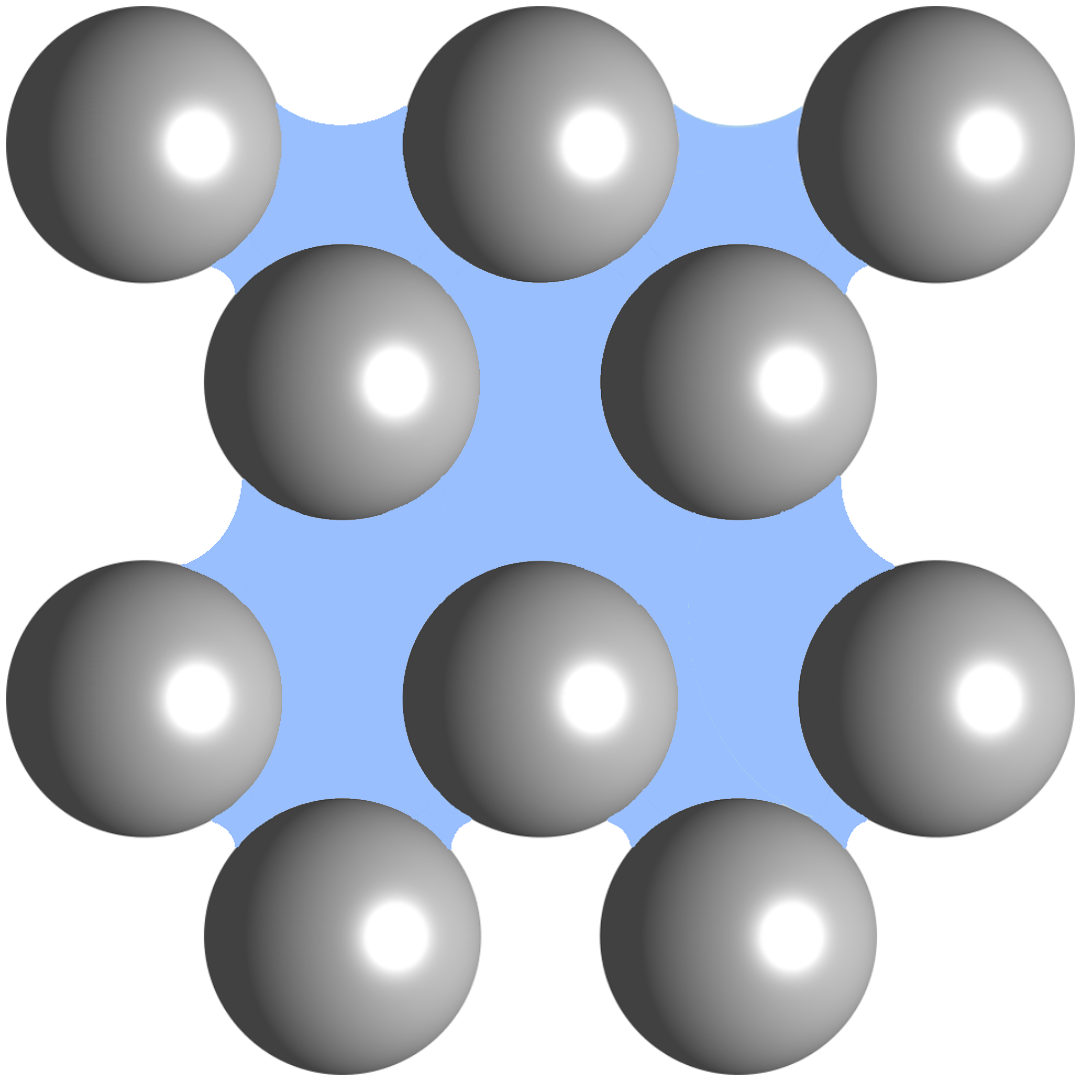}} &
		\raisebox{-.5\height}{\includegraphics[width=0.25\textwidth]{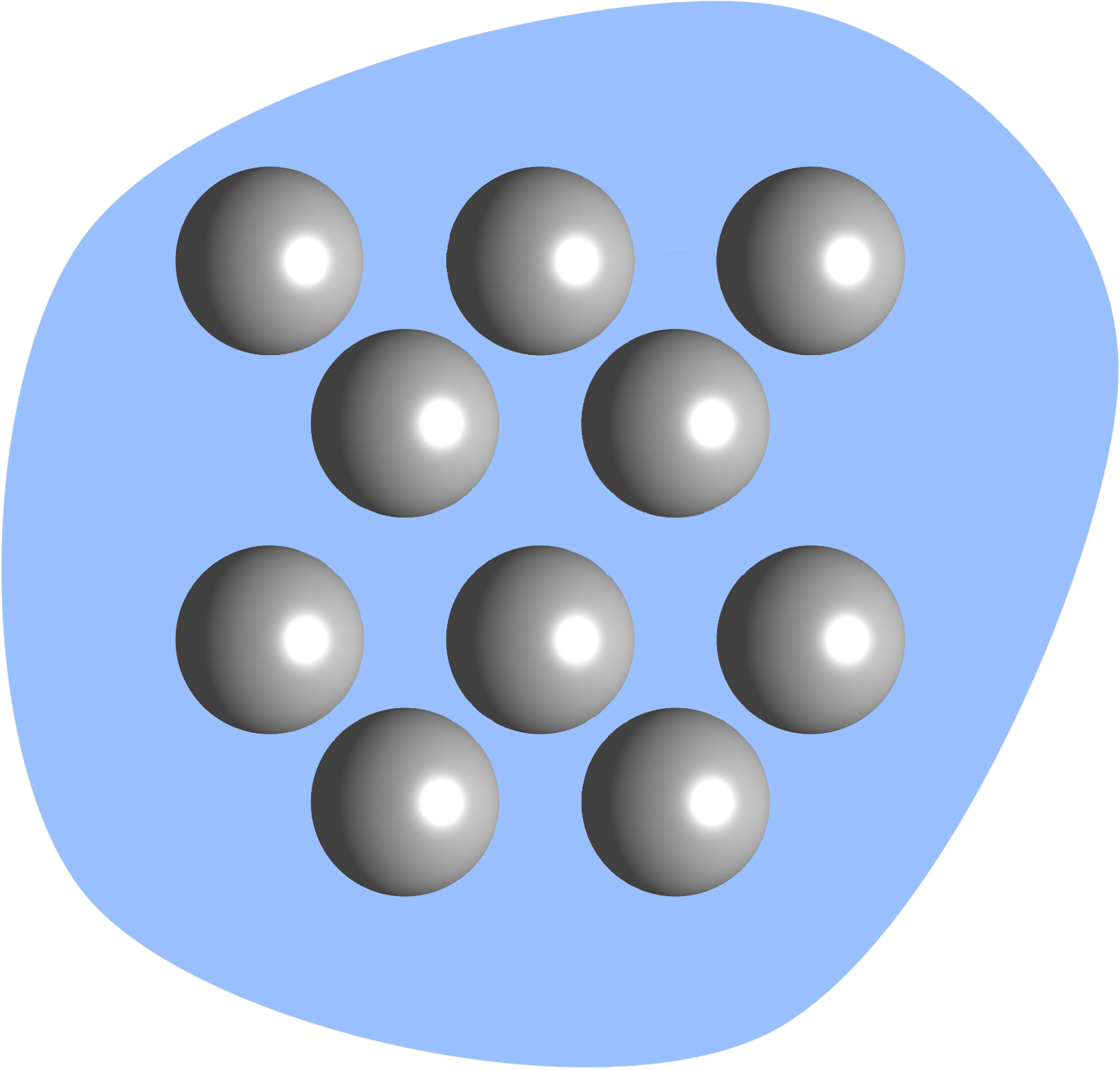}} \\
		
		\small Dry &
		\small Pendular &
		\small Funicular &
		\small Capillary &
		\small Slurry \\
	\end{tabular}
	\caption{Illustration of wetting states as a function of liquid content, adapted from \cite{mitarai2006wet}.}
	\label{fig:wet_granular_states}
\end{figure}

To model these effects accurately, it is essential to understand the physical mechanisms through which liquid affects powder flow. Among these, capillary interactions — particularly liquid bridge formation — are recognized as a primary contributor to cohesion in partially saturated systems. Herminghaus \cite{herminghaus2005dynamics} provided a foundational review on the dynamics of wet granular materials, demonstrating how capillary forces significantly affect granular stability, transport properties, and phase transitions between different wetting states. Further studies have explored how these capillary interactions affect bulk powder behavior. The classification of wet granular materials into four regimes — pendular, funicular, capillary, and slurry — based on liquid content has been well established in the literature \cite{newitt1958contribution,iveson2001nucleation,mitarai2006wet}. In the pendular state, isolated liquid bridges generate strong pairwise cohesion. With increasing liquid content, the system enters the funicular state, where bridges and partially saturated pores coexist. At still higher saturation, the capillary state forms as pores fill and cohesion is governed by suction within the liquid. In the slurry state, particles become fully immersed and capillary forces vanish. These transitions (Figure~\ref{fig:wet_granular_states}) strongly influence flowability and mechanical stability.

This study focuses on the dry-to-pendular regime, which represents the relevant moisture range for industrial HSBR powders where only isolated liquid bridges are present.

Building on this classification, several studies have examined how capillary bridges shape the flow behavior of wet granular materials in the low-moisture range relevant to this work. Althaus et al.\ \cite{althaus2012effect} showed that in the pendular state, liquid-bridge adhesion strongly controls the yield behavior of wet powders, leading to a marked increase in bulk strength. Subsequent investigations demonstrated that capillary cohesion significantly alters the velocity and solid-fraction fields, and hence the stress distribution, in sheared granular systems \cite{umer2018studies}. More recent studies have further highlighted how liquid bridges contribute to energy dissipation and force transmission during particle impacts \cite{zhang2024force} and how cohesion modifies steady-state density and rheology in both homogeneous and shear-banded flows \cite{shi2020steady}. These findings establish capillary bridging as a dominant microstructural mechanism driving mechanical response and rheology in lightly wet powders.

Beyond cohesive forces, dynamic wet granular flows are strongly shaped by the redistribution of liquid under shear. Roy et al.\ \cite{roy2017general} showed that capillary cohesion already governs local rheology in the pendular regime, while later work revealed that shearing drives moisture redistribution via a shear-rate-dependent diffusive mechanism \cite{roy2018liquid, roy2021drift}. This leads to spatial variations in local cohesion that alter bulk flow behavior. Trung Vo et al.\ \cite{trung2019agglomeration} further demonstrated that such redistribution can trigger capillary-induced agglomeration, with implications for mixing efficiency, flowability, and clogging. Collectively, these findings highlight the critical role of liquid migration and agglomeration in determining the macroscopic rheology of wet powders.

Together, these studies show that even small amounts of liquid can substantially alter the flow behavior of granular materials—through both capillary bridging and shear-induced liquid migration—underscoring the need for DEM contact models that explicitly incorporate these wetting effects.

In DEM simulations of wet granular materials, the key challenge is representing how pore liquid generates capillary forces through liquid bridge formation, rupture, and redistribution. While DEM readily captures dry particle interactions, modeling wet powders requires additional capillary force formulations that account for these liquid-mediated interactions.
Previous studies have extensively examined capillary forces in pendular liquid bridges, where individual particle pairs are connected by a small volume of liquid. Recent research has also explored funicular bridges, where liquid bridges merge across multiple particles, modifying rupture distances and cohesive forces \cite{wang2017capillary}. These findings highlight the importance of accurately modeling liquid bridge rupture criteria to ensure realistic predictions of wet powder behavior in DEM simulations. 

Two key studies have contributed to the understanding of static capillary bridge forces. Willett et al.\ \cite{willett2000capillary} provided a widely used analytical model for bridge forces between spherical particles, accounting for contact angle, separation distance, and bridge volume. Rabinovich et al.\ \cite{rabinovich2005capillary} refined this understanding by examining force–separation behavior and rupture criteria, showing through atomic force microscopy (AFM) measurements that force models depend strongly on whether bridge volume or radius is conserved. 

While these models provided valuable insight into static capillary forces, they did not account for liquid migration under dynamic flow conditions, which is critical for wet granular mixing and processing applications.
To address these limitations, Mani et al.\ \cite{mani2013liquid} extended Willett’s framework by introducing a liquid migration model that accounts for shear-induced redistribution of liquid bridges in unsaturated granular media. Their model demonstrated that under mechanical agitation, liquid is not only retained within individual bridges but also transferred between contacts, leading to heterogeneous wetting patterns that influence cohesion and agglomeration dynamics.
Building upon Willett and Mani’s foundational work, Roy et al.\ \cite{roy2018liquid} developed an advanced liquid migration model, specifically designed to simulate wet granular flows under continuous shear conditions. Roy’s model incorporates local liquid redistribution mechanisms, ensuring that liquid volume is dynamically reassigned between contacts and particle surfaces whenever a bridge ruptures or reforms. This approach refines the capillary force framework by imposing minimum liquid film and maximum liquid bridge volume constraints, preventing unbounded clustering while ensuring realistic liquid dispersion.

Although DEM provides access to particle-scale information, the contact models used in simulations are simplified representations of real particle interactions. As a result, using microscopic material properties alone does not generally reproduce the correct bulk behavior of granular systems, particularly for wet powders where liquid-mediated interactions are highly nonlinear. DEM models, therefore, require bulk-level calibration to ensure that the simulated material exhibits realistic flow behavior under relevant operating conditions.

Calibrating DEM input parameters for wet powders requires a systematic approach that ensures accurate replication of bulk material properties. Various studies have employed angle of repose (AoR) measurements \cite{roessler2018scaling, roessler2019parameter}, shear cell tests \cite{pachon2019modelling}, and uniaxial compression experiments \cite{roessler2019parameter} as calibration methods for discrete element method (DEM) parameter tuning.
One of the most widely used techniques for cohesive powder calibration is AoR-based calibration. Roessler \& Katterfeld \cite{roessler2018scaling} demonstrated that static AoR can be used to determine DEM parameters for cohesive powders by matching simulated and experimental heap formation. However, while static AoR provides insight into the stability of a granular heap, it does not fully capture the dynamic rheology of wet powders under continuous motion. Hoshishima et al.\ \cite{hoshishima2021parameter} further confirmed that relying only on static AoR for calibration can lead to inaccuracies in dynamic flow simulations, highlighting the importance of using dynamic AoR, especially for cohesive and non-spherical particles. Krantz et al.\ \cite{krantz2009characterization} reinforced this limitation by showing that static and dynamic characterization techniques can yield contradictory results, emphasizing the need for dynamic testing in flow simulations.

To address the limitations of static AoR for cohesive materials, many studies have shown that the dynamic AoR provides a more sensitive measure of flow under shear and agitation. Liu et al.\ \cite{liu2005experimental} demonstrated its dependence on operating conditions such as drum speed and fill level, while Kleinhans et al.\ \cite{kleinhans2011static} and Pourandi et al.\ \cite{pourandi2024mathematical} linked it to particle friction and the Froude number. Dynamic AoR has also been used to distinguish flow regimes in rotating drums \cite{wojtkowski2013behavior} and to quantify cohesion effects, including capillary-induced increases in surface angle \cite{jarray2019wet} and aeration-related variations with drum speed \cite{neveu2022measuring}. Although widely applied, Coetzee and Scheffler \cite{coetzee2022calibration} noted that dynamic AoR may lose sensitivity for highly cohesive powders, recommending complementary metrics for robust calibration. Collectively, these works underscore the diagnostic value of dynamic AoR and the need for DEM models capable of reproducing such moisture-dependent bulk behavior.

While several DEM contact models have attempted to represent capillary forces and liquid migration, most have been developed for ideal spherical particles and offer no practical strategy for linking the amount of liquid in a system to the capillary parameters required to reproduce wet flow in industrial powders.
This work extends an existing dry DEM contact model to wet granular materials in the pendular regime by incorporating liquid bridges that generate capillary adhesion and allowing the bridge volumes to evolve through liquid migration. The study introduces a simple schematic framework explaining how the transition from dry to wet contacts is controlled by two physically interpretable parameters—the minimum liquid film volume and the maximum liquid bridge volume. These are calibrated using dynamic AoR measurements from rotating drum experiments. By applying this procedure to two industrial powders with distinct particle size distributions and surface morphologies, we demonstrate how the calibrated model captures different wetting behaviors and identifies the conditions under which particle scaling leads to deviations at higher liquid contents. This provides a practical and experimentally anchored route for DEM-based modeling of wet industrial powders.

The paper is structured as follows. Section~\ref{sec:methods} describes the experimental and numerical methods, including particle characterization, the rotating drum setup, and the DEM framework implemented in MercuryDPM. Section~\ref{sec:results} presents the results and discussion, including the experimental AoR trends, the conceptual and simulation analysis of liquid bridge formation, and the calibration of the liquid migration parameters for both polypropylene powders. Finally, Section~\ref{sec:conclusions} summarizes the main findings and discusses their implications for modeling wet powder mixing in industrial processes.
\section{Methods} \label{sec:methods}
\subsection{Particle and liquid characterization}

This study focuses on the calibration of wet polypropylene powders to replicate real industrial conditions. Two types of powders, PP1 and PP2, were used, both produced in horizontal stirred bed reactors (HSBRs) using different catalyst systems. As a result, they differ in particle size distribution, bulk density, and surface morphology. The particle size distribution and aerated bulk density of PP1 and PP2 were characterized experimentally, as described in Reference~\cite{pourandi2024mathematical}. Laser diffraction analysis was used to determine the particle size distribution. The results are provided in Table~\ref{tab:psd}. PP2 has a larger mean particle size (909.3 $\mu$m) and a lower bulk density (368 kg/m$^3$), while PP1 has a smaller mean particle size (705.6 $\mu$m) and a higher bulk density (393 kg/m$^3$). These differences can impact the powder's response to liquid addition, as smaller and denser particles agglomerate more readily, whereas larger, more porous particles can absorb liquid into their internal structure, delaying bridge formation and reducing cohesion at low moisture levels. The mechanical properties, including elastic modulus, shear modulus, and coefficient of restitution, were obtained from material databases and literature sources and are summarized in Table~\ref{tab:material_properties}. Detailed procedures are available in Reference~\cite{pourandi2024mathematical}. The Scanning Electron Microscope (SEM) images of PP1 and PP2 illustrate their morphological differences, with PP2 displaying a more porous surface, which may influence cohesion and liquid bridge formation during wet powder mixing.

\begin{table}[H]
	\centering
	\caption{Particle Size Distribution of PP1 and PP2.}
	\label{tab:psd}
	\begin{tabular}{llllll}
		\hline
		\textbf{Sample} & \textbf{D10 ($\mu$m)} & \textbf{D25 ($\mu$m)} & \textbf{D50 ($\mu$m)} & \textbf{D75 ($\mu$m)} & \textbf{D90 ($\mu$m)}\\
		\hline
		PP1  & 304.0  & 555.5  & 736.7  & 891.6  & 1033\\
		PP2  & 506.0  & 712.7  & 904.6  & 1123  & 1345\\
		\hline
	\end{tabular}
\end{table}

\begin{table}[H]
	\centering
	\caption{Material Properties of PP1 and PP2.}
	\label{tab:material_properties}
	\begin{tabular}{lll}
		\hline
		\textbf{Parameter} & \textbf{PP1} & \textbf{PP2} \\
		\hline
		Elastic modulus $E$ (MPa)                          & 1325  & 1325 \\
		Shear modulus $G$ (MPa)                            & 400   & 400 \\
		Coefficient of restitution $e$                     & 0.5     & 0.5  \\
		Aerated (loose) bulk density $\rho$ (kg/m $^{3}$)  & 393     & 368 \\
		\hline
	\end{tabular}
\end{table}

The liquid used in this study is isopropyl alcohol (IPA), which was chosen as a substitute for propylene quench liquid because propylene is naturally in a gaseous state at ambient conditions \cite{van2024flow}. The physical properties of IPA, particularly surface tension and contact angle, are critical in determining liquid bridge formation and cohesive interactions in wet polypropylene (PP) powders. The key values, obtained from thermophysical property databases \cite{ddbstdatabase}, are shown in Table~\ref{tab:ipa_properties}.  

\begin{table}[H]
	\centering
	\caption{Physical Properties of Isopropyl Alcohol (IPA) at a temperature of 25°C.}
	\label{tab:ipa_properties}
	\begin{tabular}{ll}
		\hline
		\textbf{Parameter} & \textbf{Value} \\
		\hline
		Surface tension $\gamma$  & 21.7\,mN/m \\
		Contact angle $\theta$  & 30° \\
		\hline
	\end{tabular}
\end{table}

\subsection{Experiment}

Experiments were conducted to acquire dynamic AoR data, which is a key parameter for Discrete Element Method (DEM) calibration of wet polypropylene (PP) powders. The experimental methodology follows the approach described in Reference~\cite{pourandi2024mathematical}. We used a rotating drum with a diameter of 14 cm and a length of 18 cm, with transparent polycarbonate sidewalls allowing for direct observation of the powder flow. A high-speed camera was positioned to capture images of the powder motion, while an LED panel provided uniform illumination to enhance contrast for image analysis.

To obtain experimental AoR data for DEM calibration, polypropylene powders were pre-wetted with isopropyl alcohol (IPA) before being introduced into the drum. The samples were prepared with 0\%, 1.25\%, 2.5\%, 5\%, and 10\% liquid content (defined as the ratio $V_L / V_S$, where $V_L$ is the total liquid volume and $V_S$ is the total solid volume), ensuring a controlled variation in moisture levels. The drum was operated at a constant rotation speed of 5\,rpm. 

The Froude number \cite{juarez2011transition, jarray2019wet}, which characterizes the ratio of inertial to gravitational forces, is defined as:

\begin{equation}
	Fr = \sqrt{\frac{D\omega^2}{2g}}
\end{equation}

Here, \( D \) is the drum diameter, \( \omega \) is the drum rotational speed and \( g \) is the gravitational acceleration. For the conditions used here, the Froude number was calculated as \( Fr = 0.04 \), corresponding to the rolling flow regime observed in the experiments.

The dynamic AoR was obtained using image processing, following the methodology described in Reference~\cite{pourandi2024mathematical}. The AoR was defined as the angle between a vertical reference line and the line connecting the drum center and the center of mass of the powder bed, which was estimated from grayscale image analysis. The resulting AoR values were used to calibrate the DEM contact model parameters.

\subsection{Simulation}

\subsubsection{DEM model}

Discrete Element Method (DEM) \cite{cundall1979discrete} simulations were conducted using MercuryDPM \cite{weinhart2020fast, Thornton2023} to model the wet granular flow of polypropylene powders. The interaction between particles was captured using a contact model incorporating mechanical contact and capillary forces. Mechanical interactions were captured using the Hertz–Mindlin model \cite{di2004comparison} with added rolling resistance \cite{luding2008cohesive}, following the same implementation as in Reference~\cite{pourandi2024mathematical}. This model was employed to calculate the normal and tangential forces acting between contacting particles, considering their elastic, damping, and frictional properties. The rolling stiffness was defined to be equal to the tangential stiffness \cite{luding2008cohesive}.

Within the Hertz--Mindlin contact model, the sliding ($\mu_s$) and rolling ($\mu_r$) friction coefficients control dissipation and resistance to motion at particle contacts. For PP2, the values were taken directly from Reference~\cite{pourandi2024mathematical}, where the dry contact model was calibrated using the same polypropylene powder as in this study. For PP1, the same experimental protocol was applied here to determine the corresponding coefficients. In all cases, the coefficients were selected such that the simulated AoR matched the experimental measurements. 

To reduce computational cost, a scaling approach was adopted in which the particle size was increased by a factor $l$, following the same strategy as in Reference~\cite{pourandi2024mathematical}. Simulating the system at the original particle scale would be prohibitively expensive due to the very large number of particles required. For example, for PP2 at a scaling factor of $l = 3$, more than $10^5$ particles (101{,}200) were required to fill the drum, whereas simulations at the original scale would require several million particles. Particle scaling also increases the numerically stable time step and, consequently, reduces the overall computational cost. The resulting values of $\mu_s$ and $\mu_r$ for both PP1 and PP2 are reported in Table~\ref{tab:friction_coefficients}.

\begin{table}[H]
	\centering
	\caption{Friction coefficients of PP1 and PP2 at different scaling factors ($l$).}
	\label{tab:friction_coefficients}
	\begin{tabular}{lccc}
		\hline
		\textbf{Parameter} & \textbf{PP1 ($l$ = 5)} & \textbf{PP2 ($l$ = 3)} & \textbf{PP2 ($l$ = 5)} \\
		\hline
		Sliding friction $\mu_s$ & 0.8 & 0.8 & 0.8 \\
		Rolling friction $\mu_r$ & 0.3 & 0.4 & 0.3 \\
		\hline
	\end{tabular}
\end{table}

The cohesive behavior of the wet particles due to the capillary bridges was modeled using the hysteretic nonlinear liquid bridge force model originally proposed by Willett et al.\ \cite{willett2000capillary} and implemented in MercuryDPM as described in Reference~\cite{roy2018liquid}. The simulation also accounted for the redistribution of liquid within the system using a liquid migration model \cite{roy2018liquid}, ensuring accurate representation of liquid transport between particle contacts. The adhesive force generated by a liquid bridge between particles $i$ and $j$ is expressed as

\begin{equation}
	f^{i,j}_c = \frac{f_c^{0}}{1 + 1.05\bar{S} + 2.5\bar{S}^2},
	%\tag{4.2a}
\end{equation}

where

\begin{equation}
	f_c^{0} = 2\pi r_{\mathrm{eff}} \gamma \cos\theta,
	%\tag{4.2b}
\end{equation}

is the maximum capillary force at contact (pull-off force), with $r_{\mathrm{eff}}$ the harmonic mean of the interacting particle radii, $\gamma$ the surface tension, and $\theta$ the contact angle. $f_c^{0}$ is independent of the bridge volume $V_b$; the $V_b$-dependence enters only through the dimensionless separation distance $\bar{S} = S\sqrt{r_{\mathrm{eff}}/V_b}$, where $S$ is the separation distance, i.e., the gap between the particle surfaces. This nondimensionalization scales $S$ by the effective particle radius and liquid bridge volume. 
Liquid bridges rupture when the separation distance reaches a critical value

\begin{equation}
	S_c = \left( 1 + \frac{\theta}{2} \right) V_b^{1/3},
\end{equation}

beyond which adhesion can no longer be sustained. 
This rupture criterion ensures that capillary forces do not persist unrealistically once particles separate beyond the physically observed limits of liquid bridges. Upon rupture, the bridge volume is redistributed among neighboring contacts, ensuring liquid conservation within the system.

To preserve physically meaningful cohesive behavior when particle size was scaled, the pressure-based Bond number framework proposed by Roy et al.\ \cite{roy2017general} and applied in the scaling strategy of Larijani et al.\ \cite{larijani2025coarse} was adopted. Rather than using the classical Bond number based on single-particle weight, this formulation relates cohesive stress to local confining stress in the granular bulk and is therefore more appropriate for dense shear-driven systems.

Following the pressure-based scaling rule derived by Larijani et al.\ \cite{larijani2025coarse}, physical similarity is achieved by preserving the ratio between capillary stress and confining stress. Accordingly, the liquid surface tension was scaled proportionally with the particle size scaling factor \(l\), such that

\begin{equation}
	\gamma_{\text{scaled}} = l \, \gamma .
\end{equation}

This scaling preserves the balance between capillary and confining stresses and therefore maintains the correct bulk rheology of the wet granular material after particle upscaling. It should be noted, however, that the size of cohesive agglomerates is not controlled by this scaling rule and may become larger than physically realistic at high liquid contents.

To realistically capture the distribution of liquid bridges in the granular bulk, a liquid migration model was implemented, as described in Reference~\cite{roy2018liquid}. In this model, liquid can be present on particle surfaces, represented by the liquid film volume $V_f^i$, as well as in liquid bridges, represented by the liquid bridge volume $V_b^{ij}$. Those bridges form and break according to the above rupture and formation criteria.

When two particles make contact ($S<0$), and if the sum of their liquid film volumes exceeds the specified minimum liquid film volume ($V_f^i + V_f^j > V_{\min}$), a portion of this liquid film is transferred to the liquid bridge that forms between them. The liquid bridge volume is defined as

\begin{equation}
	V_b^{ij} = \min \left( V_f^i + V_f^j, V_{\max} \right),
\end{equation}

where $V_f^i$ and $V_f^j$ are the available liquid films on the particles. The maximum liquid bridge volume $V_{\max}$ limits liquid coalescence, preventing unrealistically large bridges and excessive accumulation in individual contacts. In addition, a minimum liquid film volume $V_{\min}$ accounts for liquid trapped in surface asperities or within pores, so a bridge only forms when this threshold is exceeded. Following the formulation of Roy et al.\ \cite{roy2018liquid}, these two parameters govern the liquid distribution in the system and need to be calibrated.

By integrating both contact forces and capillary interactions, this DEM model provides a realistic representation of wet powder behavior, allowing for direct comparison with experimental data.

\begin{figure}[H]
	\centering
	\renewcommand{\arraystretch}{1.2}
	\begin{tabular}{c@{\hspace{12pt}}c}
		\raisebox{-.5\height}{\includegraphics[width=0.35\textwidth]{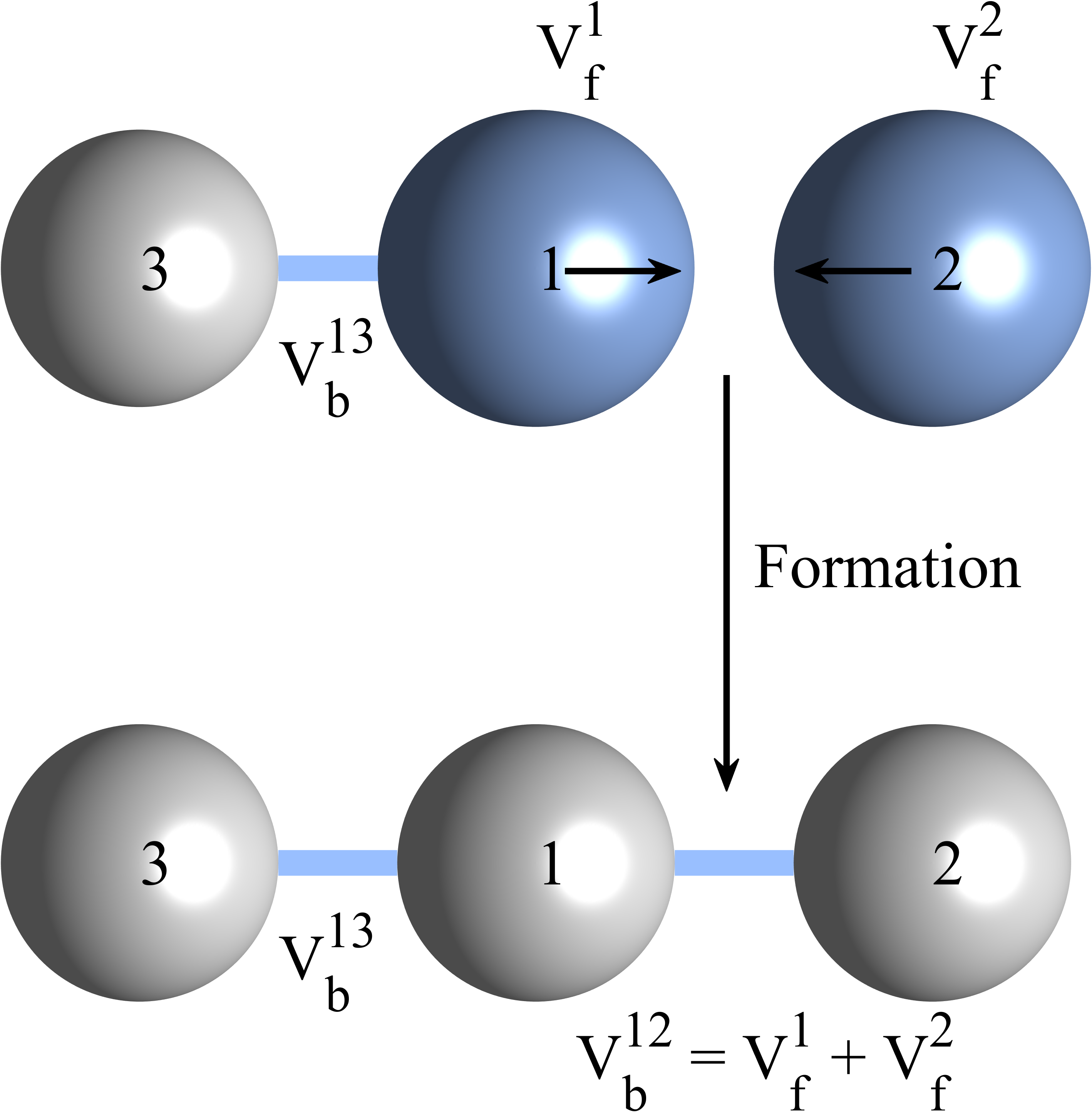}} &
		\raisebox{-.5\height}{\includegraphics[width=0.44\textwidth]{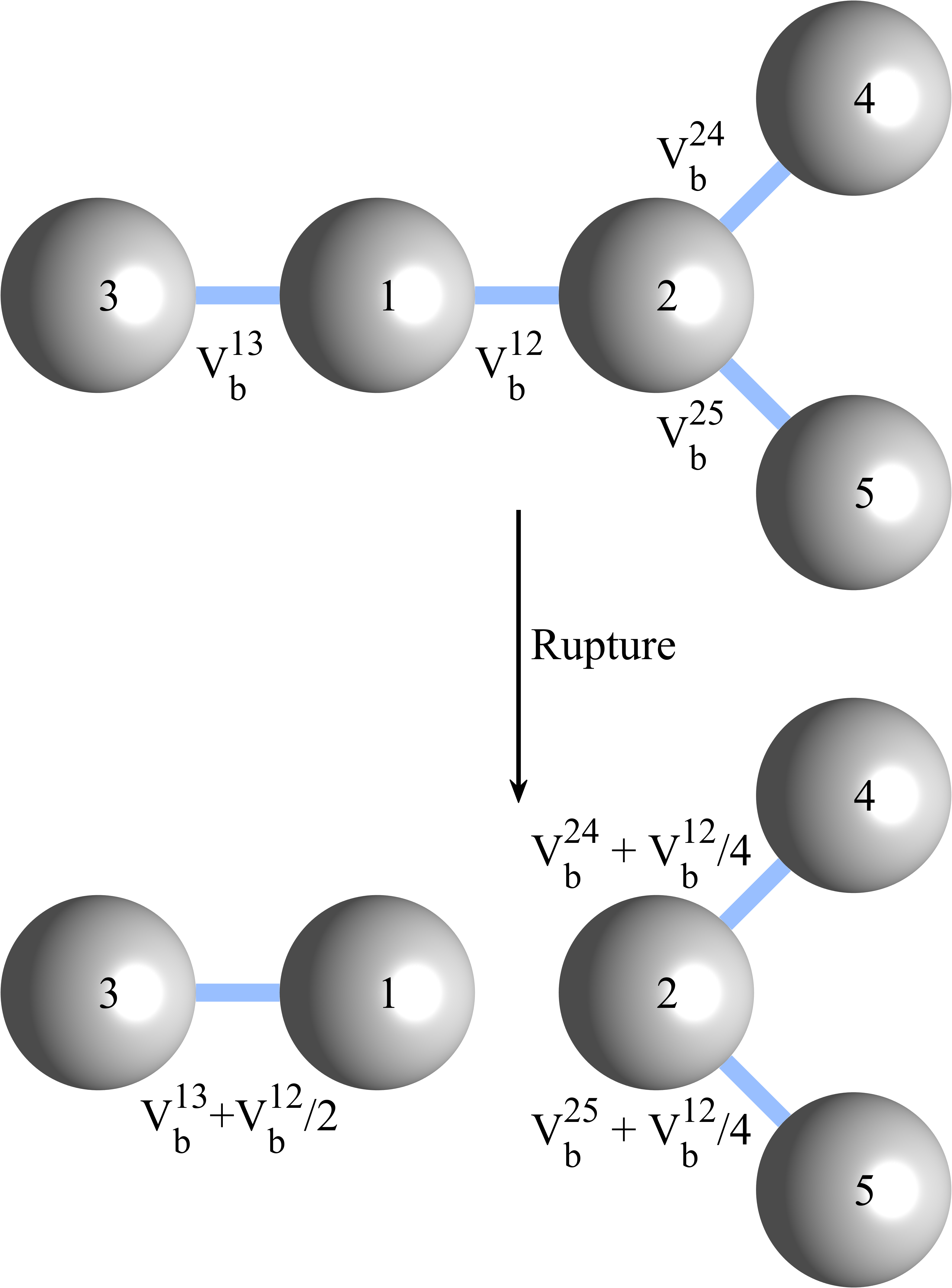}} \\
	\end{tabular}
	\caption{Illustration of the formation (left) and rupture (right) of a liquid bridge. Blue particles are wetted ($V_f > 0$), gray particles are dry ($V_f = 0$), and blue lines represent liquid bridges.}
	\label{fig:LiquidBridgeMechanism}
\end{figure}

\subsubsection{Simulation procedure}

The aim of these simulations was to calibrate \(V_{\min}\) and \(V_{\max}\) in the liquid migration model to match experimental dynamic AoR values.
Simulations were conducted to replicate the experimental conditions, using the experimental rotating drum system where polypropylene powders were subjected to varying liquid contents \(V_L/V_S\), as shown in Table~\ref{tab:PP2CalibrationParameters}.
The Froude number (\( Fr \)) was preserved by using the same drum diameter and rotational speed as in the experiments, ensuring dynamic similarity between the experimental and simulated systems.

First, simulations were performed for PP2, using a scaling factor of \(l = 5\).
For PP2, \(V_{\min}/\bar{V}_S\) was fixed at 10\%.
The maximum liquid bridge volume (\(V_{\max}\)) was varied from 0.1\% to 10\% of the mean solid volume (\(\bar{V}_S = V_S / N_p\)), as shown in Table~\ref{tab:PP2CalibrationParameters}. The simulations were repeated for each liquid content at which experiments were taken.

\begin{table}[H]
	\centering
	\caption{Parameter values used in the calibration study for PP2 ($l = 3$ and $l = 5$).}
	\label{tab:PP2CalibrationParameters}
	\begin{tabular}{ll}
		\hline
		\textbf{Parameter} & \textbf{Values} \\
		\hline
		\(V_L/V_S\) (\%)  & 0.0, 1.25, 2.5, 5.0, 10.0 \\
		\(V_{\max}/\bar{V}_S\) (\%) & 0.1, 0.25, 1.0, 4.0, 10.0 \\
		\hline
	\end{tabular}
\end{table}

Second, simulations were conducted for PP1. The maximum liquid bridge volume (\(V_{\max}\)) was again varied from 0.1\% to 10\% of the mean solid volume (\(\bar{V}_S\)) while maintaining the minimum liquid film volume (\(V_{\min}\)) at zero. These simulations were then repeated with three other chosen \(V_{\min}\), as summarized in Table~\ref{tab:PP1CalibrationParameters}.

\begin{table}[H]
	\centering
	\caption{Parameter values used in the calibration study for PP1 ($l = 5$).}
	\label{tab:PP1CalibrationParameters}
	\begin{tabular}{ll}
		\hline
		\textbf{Parameter} & \textbf{Values} \\
		\hline
		\(V_L/V_S\) (\%) & 0.0, 1.25, 2.5, 5.0, 10.0 \\
		\(V_{\min}/\bar{V}_S\) (\%) & 0.0, 0.625, 1.25, 2.25 \\
		\(V_{\max}/\bar{V}_S\) (\%) & 0.1, 0.25, 1.0, 4.0, 10.0 \\
		\hline
	\end{tabular}
\end{table}

Consistent with the experimental procedure, the dynamic AoR in the simulations was determined by calculating the angle between the vertical reference axis and the line connecting the drum center and the center of mass of the powder bed. The center of mass was computed directly from particle positions at each sampling instant, and the AoR was extracted from its angular position relative to gravity direction. This ensured consistent definition of the dynamic AoR between experiments and simulations, allowing for direct comparison. The dynamic AoR was evaluated under steady-state conditions, and the reported values therefore represent the time-averaged AoR in the quasi-steady state.

To ensure numerical stability, the simulation time step (\(\Delta t\)) was set to 20\% of the Rayleigh time step (\(\Delta t_{\mathrm{Rayleigh}}\)); see Reference~\cite{pourandi2024mathematical} for its definition and calculation procedure. To reduce computation time, the particle stiffness was reduced by a factor of \(10^3\), following the approach in References~\cite{yan2015discrete, lommen2014speedup}. This reduction did not affect the numerical stability or the behavior of the simulated particle flow.
\section{Results and discussion}\label{sec:results}
As mentioned, the simulation setup was designed to replicate the experimental system. We will now calibrate \(V_{\min}\) and \(V_{\max}\) so that the simulated dynamic AoR matches the experimental results.

\subsection{Experimental results}

\begin{figure}[H]
	\centering
	\includegraphics[width=0.6\textwidth]{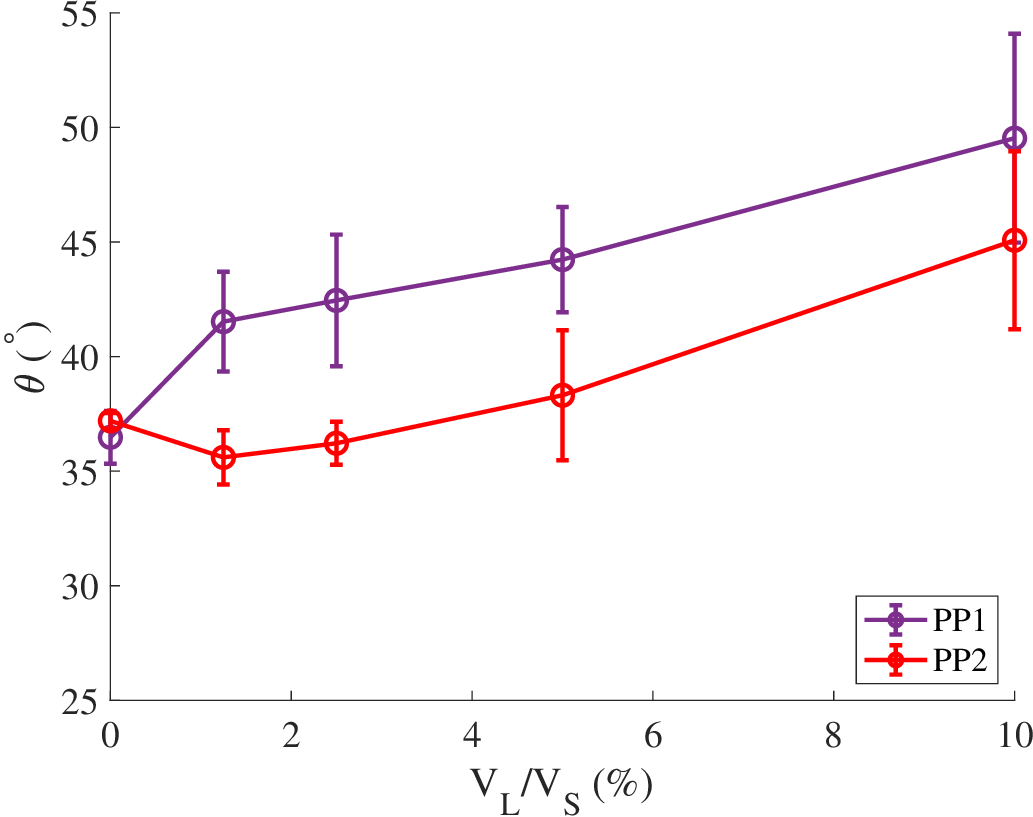}
	\caption{Experimental dynamic angle of repose ($\theta$) for PP1 and PP2 as a function of liquid content.}
	\label{fig:ExperimentalAoR}
\end{figure}

Figure~\ref{fig:ExperimentalAoR} presents the experimentally measured AoR values for both PP1 and PP2. In both cases, the AoR is at its lowest in the dry condition and increases as the liquid volume increases. The standard deviation of AoR measurements provides further insight into material behavior. A high standard deviation suggests that the material exhibits cohesive behavior, forming agglomerates that move intermittently. In contrast, a low standard deviation indicates that the material remains non-cohesive, flowing smoothly without significant aggregation.

A distinct difference in cohesive behavior is observed between PP1 and PP2. PP2 remains non-cohesive at lower liquid concentrations (\(\leq5\)\,vol\%), only transitioning to a cohesive state at higher liquid contents (\(>5\)\,vol\%). In contrast, PP1 exhibits cohesive behavior immediately when any liquid is added (\(\geq1.25\)\,vol\%). 

\begin{figure}[H]
	\centering
	\renewcommand{\arraystretch}{1.2}
	\begin{tabular}{c@{\hspace{2pt}}c@{\hspace{2pt}}c@{\hspace{2pt}}c@{\hspace{2pt}}c@{\hspace{2pt}}c}
		\raisebox{0.5\height}{\small PP1} &
		\includegraphics[width=0.18\textwidth]{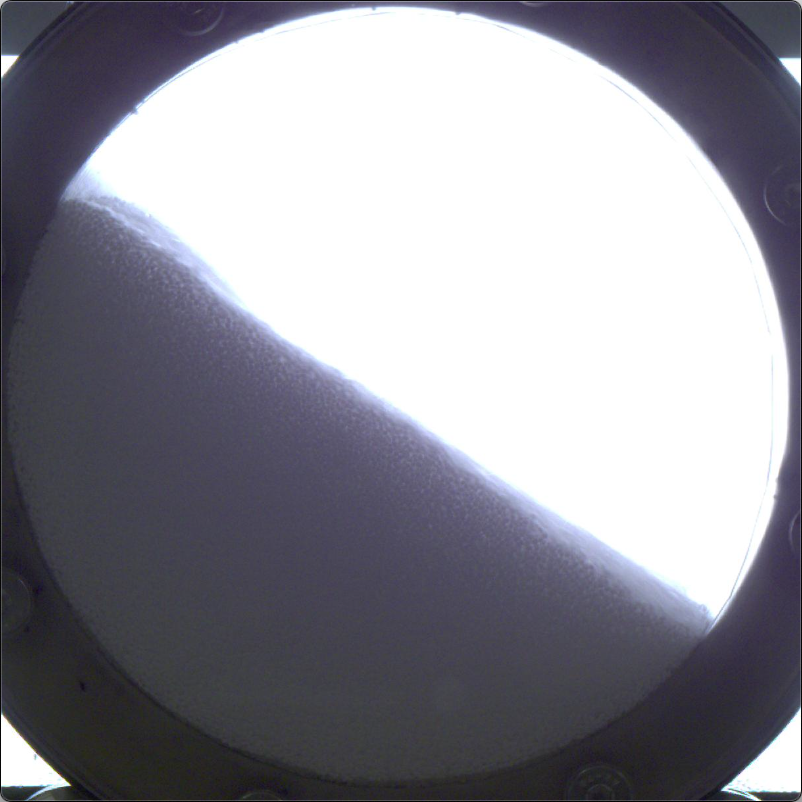} &
		\includegraphics[width=0.18\textwidth]{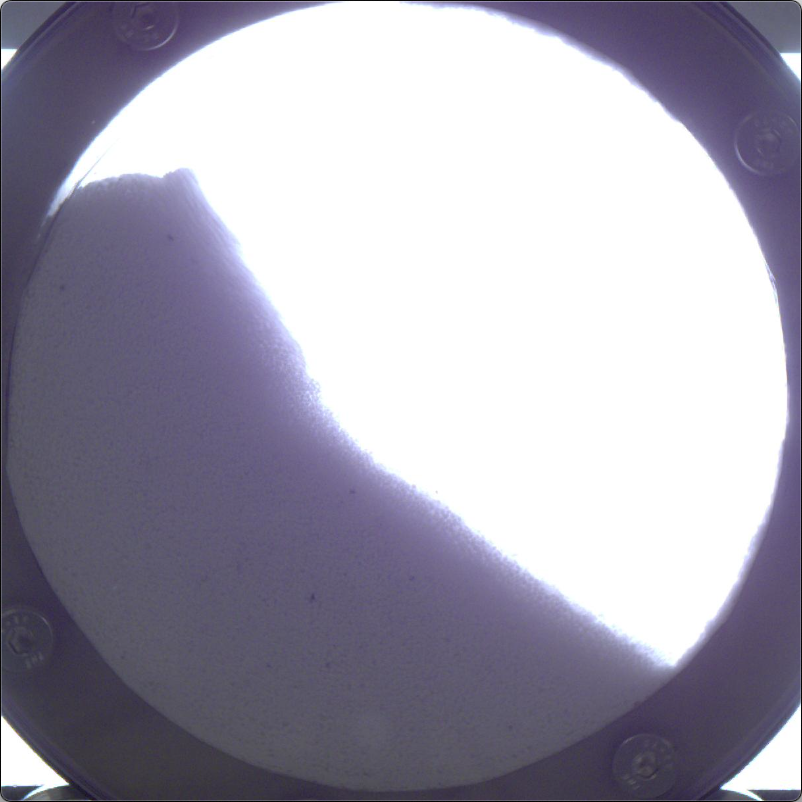} &
		\includegraphics[width=0.18\textwidth]{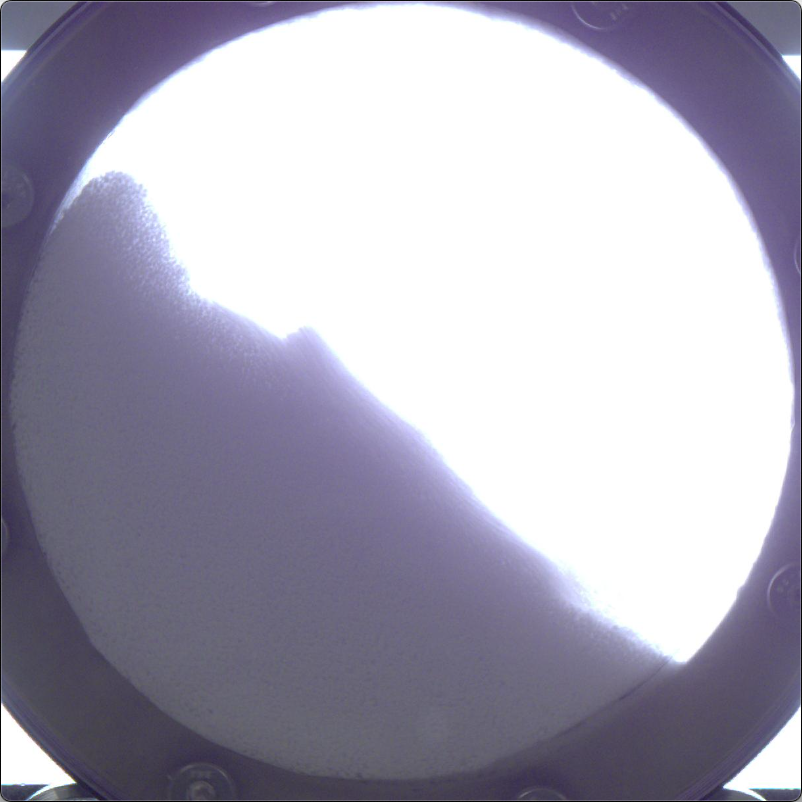} &
		\includegraphics[width=0.18\textwidth]{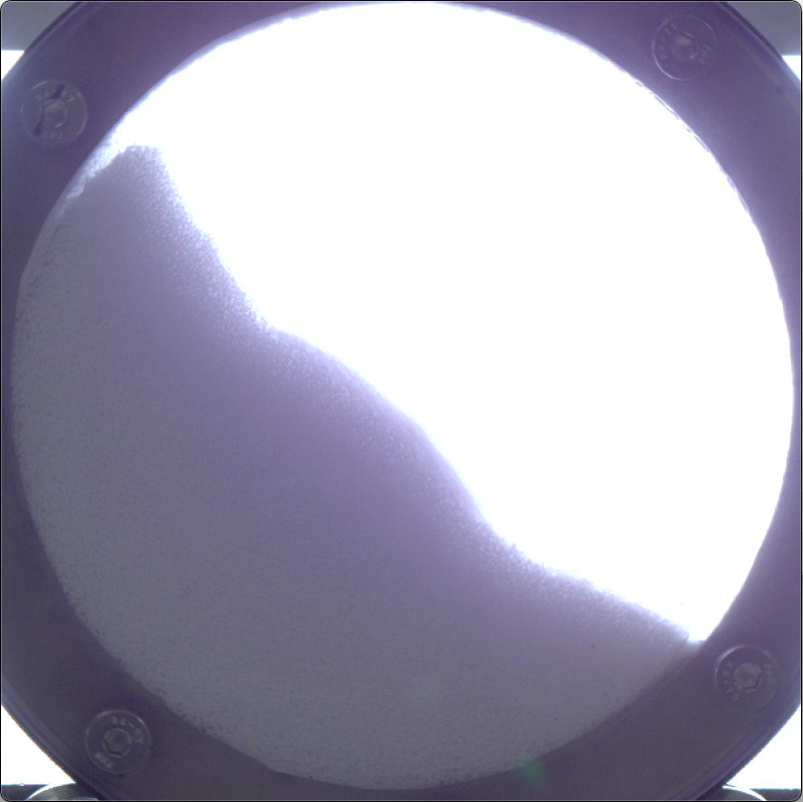} &
		\includegraphics[width=0.18\textwidth]{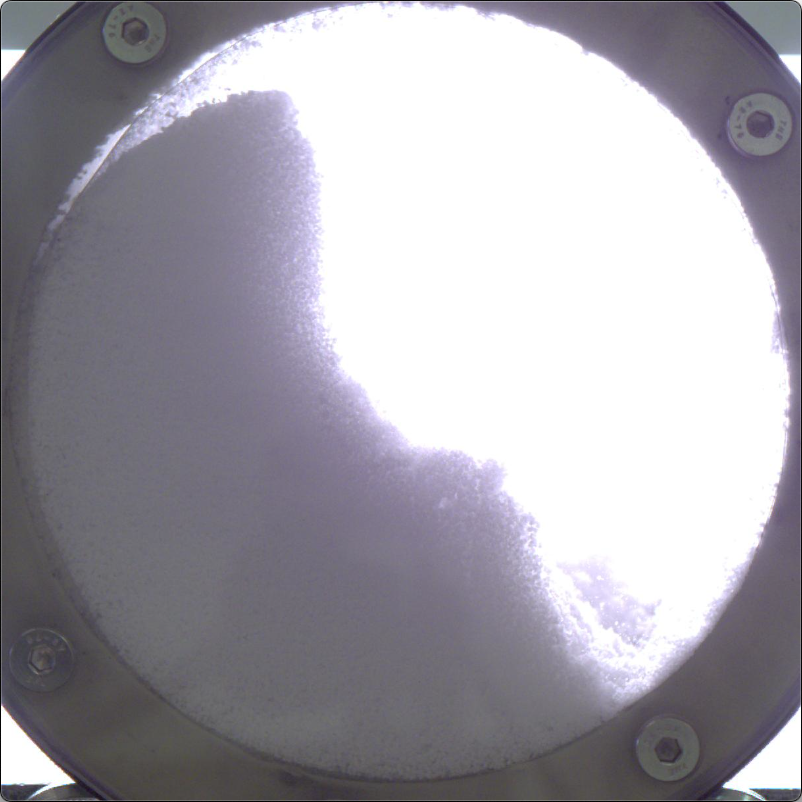} \\
		
		\raisebox{0.5\height}{\small PP2} &
		\includegraphics[width=0.18\textwidth]{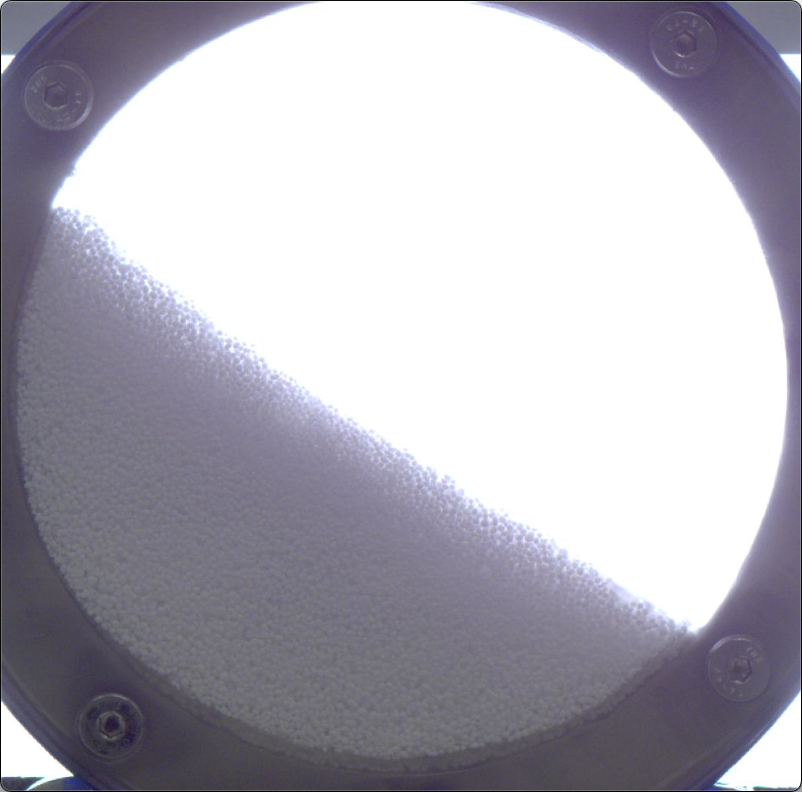} &
		\includegraphics[width=0.18\textwidth]{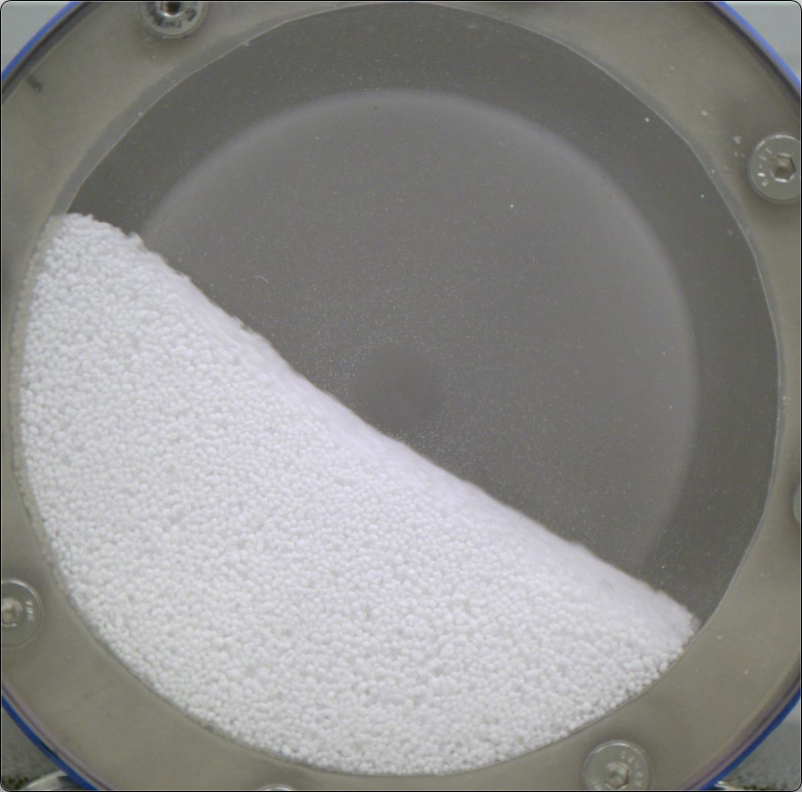} &
		\includegraphics[width=0.18\textwidth]{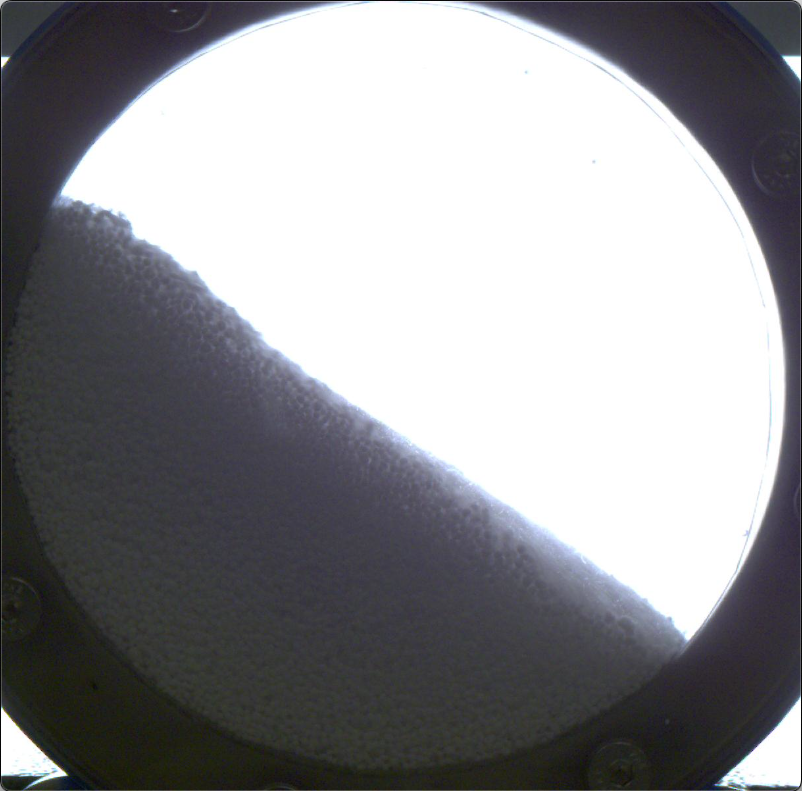} &
		\includegraphics[width=0.18\textwidth]{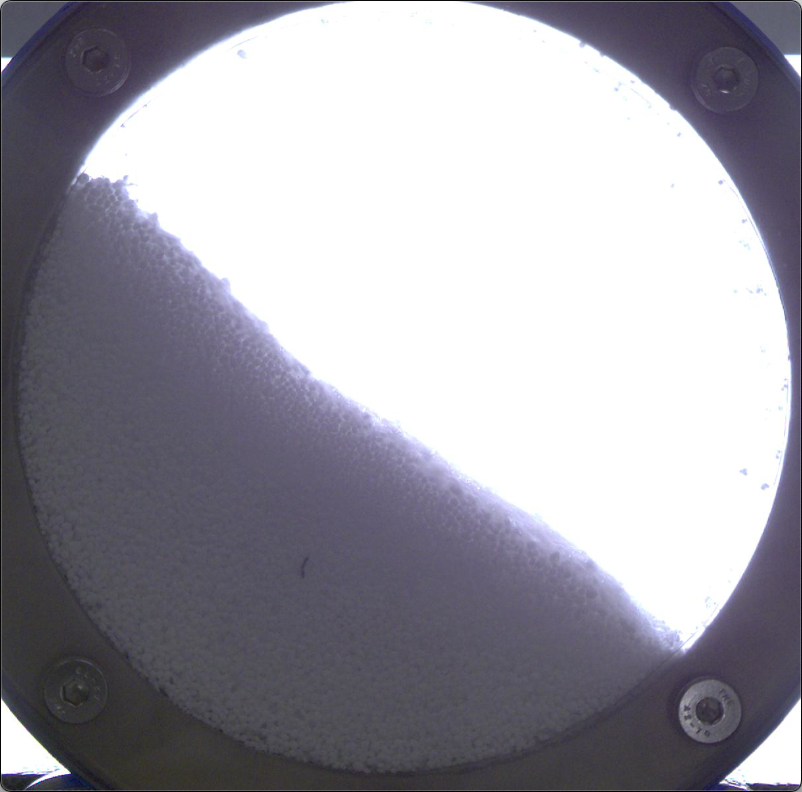} &
		\includegraphics[width=0.18\textwidth]{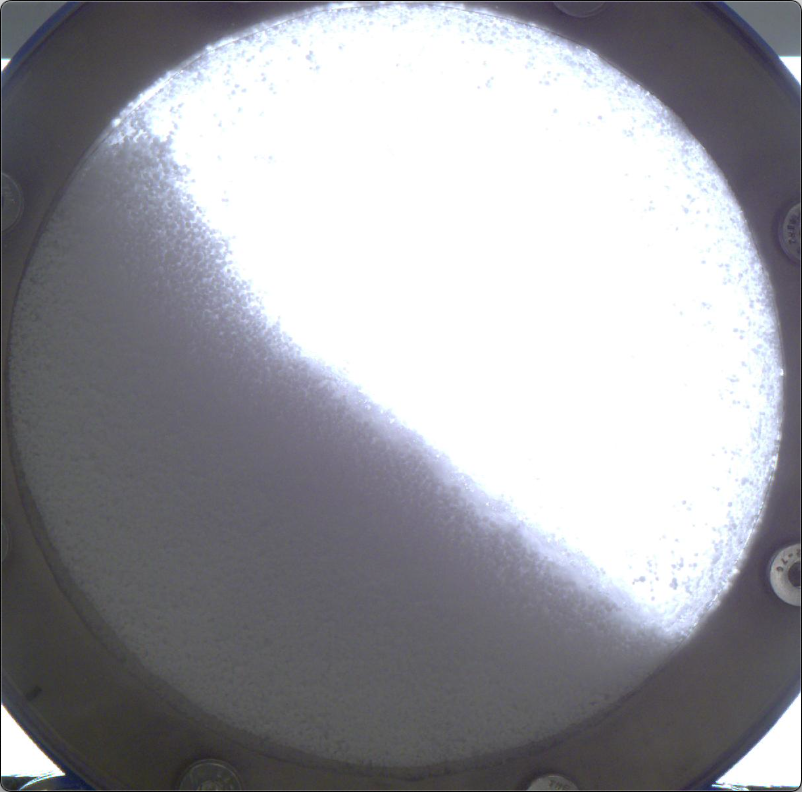} \\
		
		& \small 0 vol\% &
		\small 1.25 vol\% &
		\small 2.5 vol\% &
		\small 5 vol\% &
		\small 10 vol\% \\
	\end{tabular}
	\caption{Experimental snapshots of PP1 and PP2 at different liquid contents, showing the flow behavior.}
	\label{fig:PP1_PP2_snapshots}
\end{figure}

These trends are further supported by experimental images (Figure~\ref{fig:PP1_PP2_snapshots}), which visually capture the differences in flow behavior between PP1 and PP2 at various liquid contents. In PP2, the material remains free-flowing up to 5\,vol\% liquid content, with cohesive behavior only becoming apparent at higher liquid contents. In contrast, PP1 exhibits cohesive flow behavior almost immediately upon the addition of liquid. This difference is attributed to the higher porosity of PP2, which allows it to absorb more liquid into its internal pore structure before liquid bridges form at particle contacts. Consequently, the onset of cohesion in PP2 occurs at higher liquid contents compared to PP1. These visual observations are consistent with the measured AoR values and further confirm the distinct wetting behaviors of the two powders.

\subsection{Conceptual framework and simulation insight}

\begin{figure}[H]
	\centering
	\includegraphics[width=0.6\textwidth]{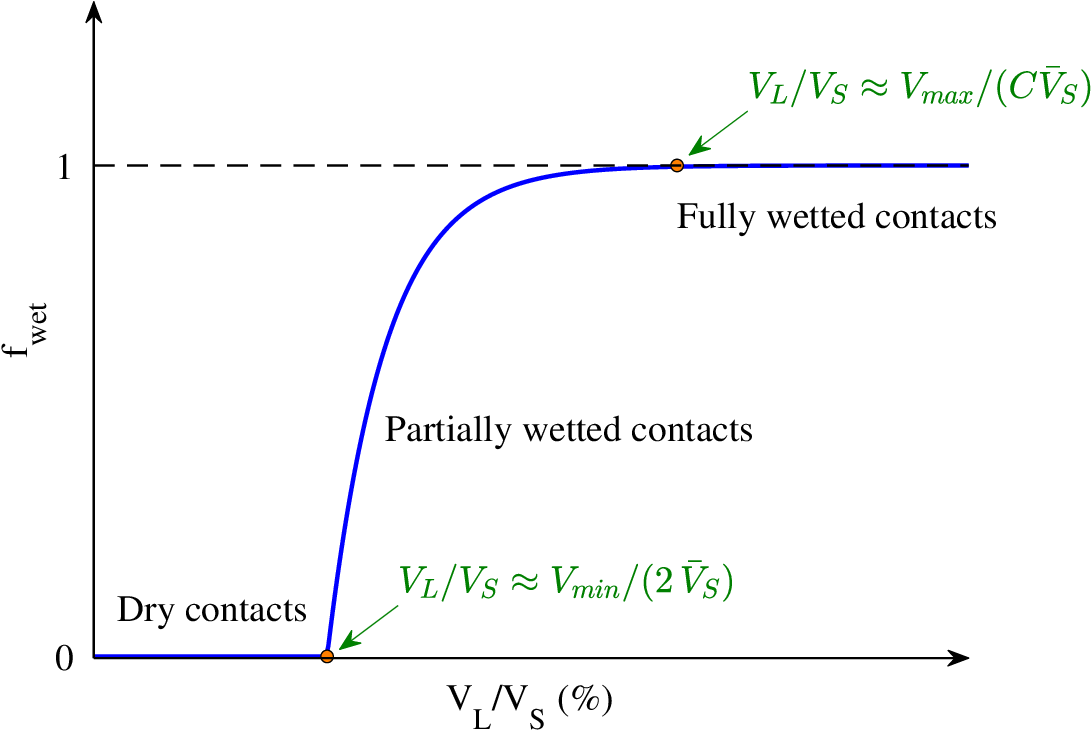}
	\caption{Schematic relationship between the fraction of wet contacts ($f_{\mathrm{wet}}$) and liquid content, highlighting the influence of $V_{\min}$ and $V_{\max}$.}
	\label{fig:Schematic_AoR}
\end{figure}

The relationship between liquid content and cohesion can be understood from the schematic shown in Figure~\ref{fig:Schematic_AoR}, which illustrates the fraction of wet contacts, denoted as $f_{\mathrm{wet}}$ (i.e., the normalized number of liquid bridges), as a function of liquid content. This parameter is defined as the ratio between the number of wet contacts (those connected by a liquid bridge) and the total number of (wet and dry) particle contacts. It therefore provides a quantitative measure of how many particle interactions are affected by capillary forces.

At very low liquid contents, all contacts are dry and $f_{\mathrm{wet}}$ is close to zero. Once the liquid volume exceeds the threshold defined by \(V_{\min}\), liquid bridges begin to form between particles, and $f_{\mathrm{wet}}$ increases sharply, marking the onset of cohesion. As the liquid content continues to rise, the number of wet contacts approaches a maximum and eventually saturates when the bridge capacity, characterized by \(V_{\max}\), is reached. Beyond this point, additional liquid no longer increases the number of bridges but instead enlarges the existing ones.

This framework explains the experimentally observed difference between PP1 and PP2. The delayed onset of cohesion in PP2 can be attributed to its higher porosity, which allows liquid to be absorbed into internal pore structures before contributing to liquid bridge formation at particle contacts. As a result, a larger fraction of the added liquid in PP2 remains unavailable for bridge formation at low liquid contents, delaying the increase in the number of wet contacts. In contrast, PP1 exhibits a more rapid increase in liquid bridges with liquid addition, consistent with earlier onset of cohesive behavior. In the model, this behavior is captured through the parameters \(V_{\min}\) and \(V_{\max}\), which control the onset and saturation of bridge formation, respectively.

\begin{figure}[H]
	\centering
	\includegraphics[width=0.6\textwidth]{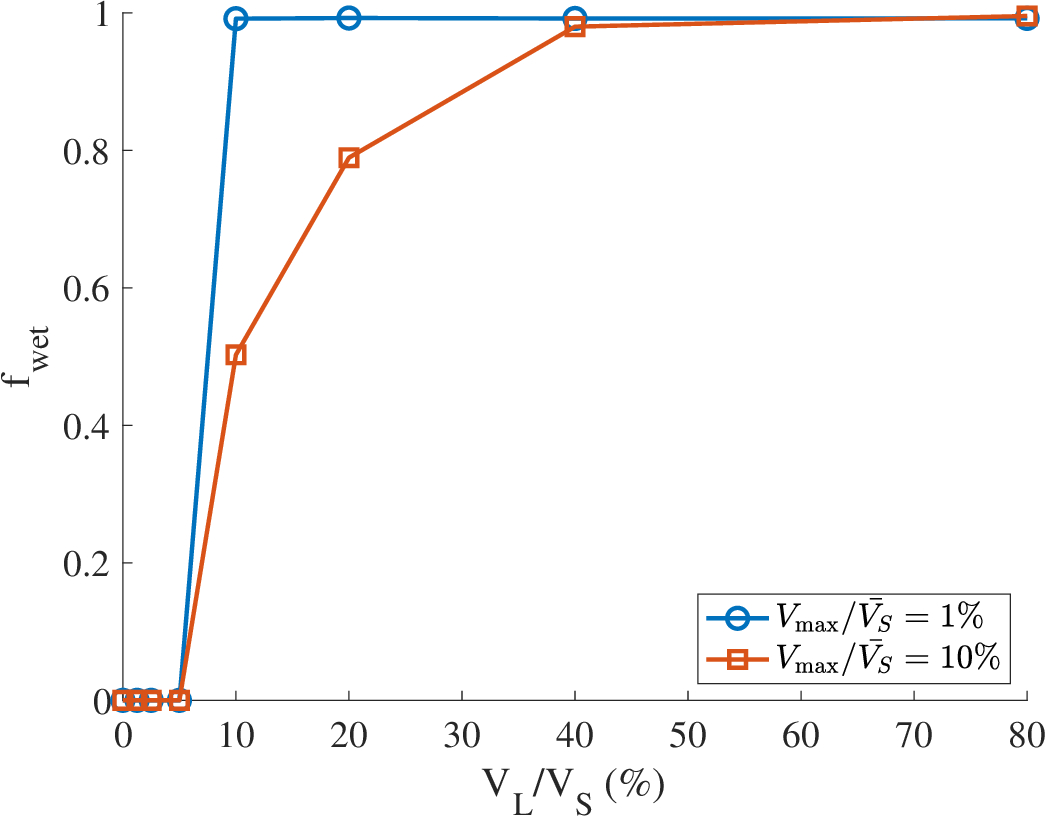}
	\caption{Fraction of wet contacts ($f_{\mathrm{wet}}$) for PP2 ($l = 5$) as a function of liquid content, comparing two values of maximum liquid bridge volume.}
	\label{fig:Fraction_vs_LVSV}
\end{figure}

This conceptual trend is confirmed by the simulation results presented in Figure~\ref{fig:Fraction_vs_LVSV}, which show the fraction of wet contacts, $f_{\mathrm{wet}}$, obtained numerically for two values of \(V_{\max}\). For \(V_{\max}/\bar{V}_S = 1\%\), $f_{\mathrm{wet}}$ increases rapidly and reaches saturation at relatively low liquid contents (around 10\,vol\%). In contrast, for \(V_{\max}/\bar{V}_S = 10\%\), $f_{\mathrm{wet}}$ increases more gradually, with saturation occurring at much higher liquid contents (40--80\,vol\%).

This difference reflects the larger capacity of liquid bridges at higher \(V_{\max}\) values: when each bridge can accommodate more liquid, a greater amount of liquid must be added before all contacts become wetted, delaying the point of full saturation. Consequently, systems with higher \(V_{\max}\) sustain cohesive growth over a wider range of liquid contents and exhibit stronger capillary effects at higher moisture levels.

Together, Figures~\ref{fig:Schematic_AoR} and~\ref{fig:Fraction_vs_LVSV} demonstrate the close correspondence between the conceptual model and the DEM simulations. The comparison confirms that \(V_{\min}\) governs the onset of bridge formation, while \(V_{\max}\) controls the saturation of cohesive interactions, providing a clear physical basis for calibrating these parameters in subsequent analyses. 

Although not shown here, the total number of wet contacts increases with liquid content for each fixed \(V_{\max}\). For larger \(V_{\max}\), saturation is delayed to higher liquid contents because each bridge can hold more liquid.

\subsection{Calibration of PP2}

The calibration of PP2 focused on identifying the liquid bridge parameters \(V_{\min}\) and \(V_{\max}\) that best reproduce the experimentally observed transition from non-cohesive to cohesive behavior, as reflected in the dynamic AoR.

\begin{figure}[H]
	\centering
	\includegraphics[width=0.6\textwidth]{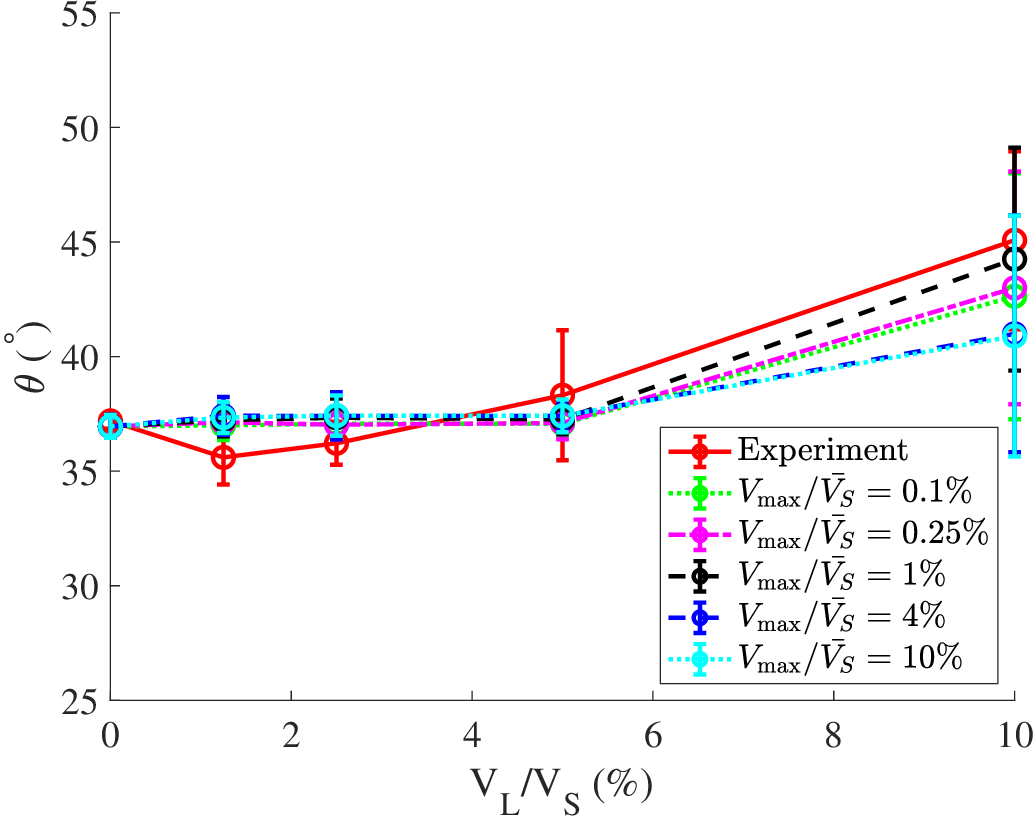}
	\caption{Experimental and simulated dynamic angle of repose (\(\theta\)) for PP2 at scaling factor \(l = 5\), comparing different values of maximum liquid bridge volume to determine the best match with experiments.}
	\label{fig:SimulationAndExperimentalPP2AOR_l5}
\end{figure}

The parameter \(V_{\min}\) was determined directly from the experimental observations. As shown in Figure~\ref{fig:SimulationAndExperimentalPP2AOR_l5}, the transition from free-flowing to cohesive behavior for PP2 occurs at approximately 5\,vol\% liquid content. According to the schematic framework in Figure~\ref{fig:Schematic_AoR}, a liquid bridge forms when the combined film volume at a contact exceeds \(V_{\min}\). Thus, the transition near 5\,vol\% corresponds to \(V_{\min}/\bar V_S \approx 10\%\).

The parameter \(V_{\max}\) was then identified by comparing simulated AoR results obtained for several \(V_{\max}\) values with the experimental measurements as illustrated in Figure~\ref{fig:SimulationAndExperimentalPP2AOR_l5}. The case that provided the closest match to the experimental AoR was selected as the calibrated value. As shown in Figure~\ref{fig:SimulationAndExperimentalPP2AOR_l5}, the simulation reproduced the experimental trend most accurately for \(V_{\max}/\bar{V}_S = 1\%\), confirming that the chosen parameters (\(V_{\min}/\bar{V}_S = 10\%\), \(V_{\max}/\bar{V}_S = 1\%\)) reliably capture both the non-cohesive regime at low liquid contents (0--5~vol\%) and the rapid increase in AoR once liquid bridges form.

\begin{figure}[H]
	\centering
	\includegraphics[width=0.6\textwidth]{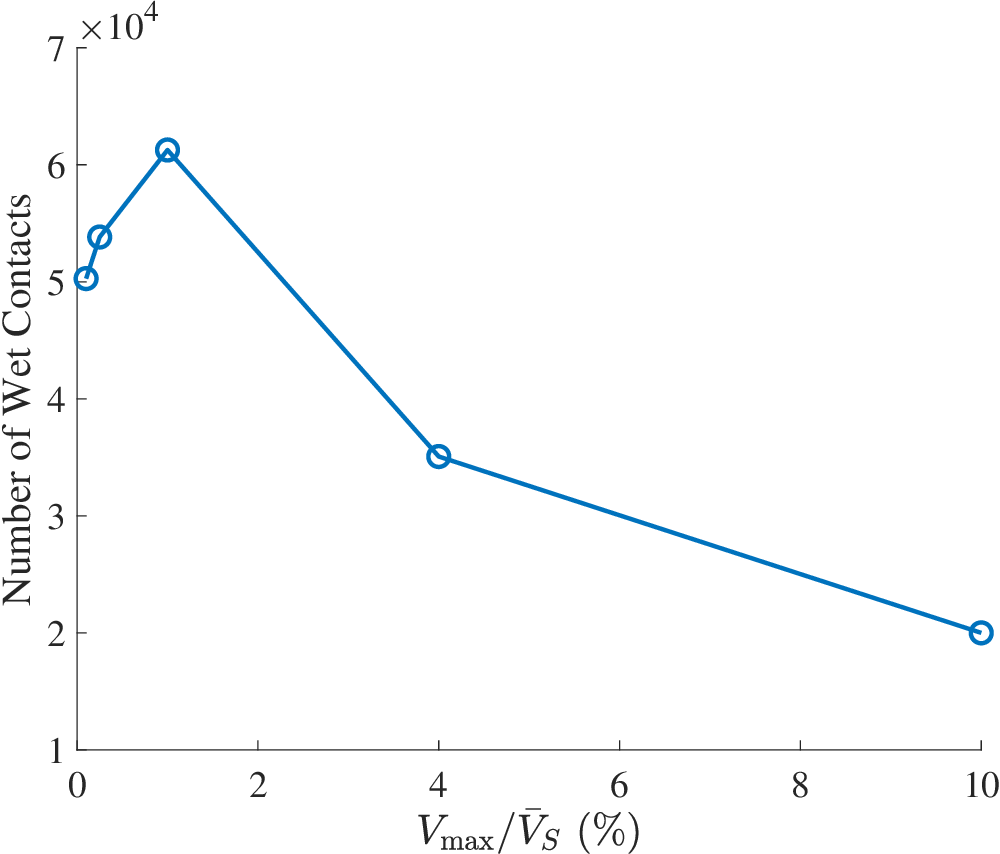}
	\caption{Total number of wet contacts for PP2 ($l = 5$) at a liquid content of 10~vol\% as a function of maximum liquid bridge volume.}
	\label{fig:PP2_WetContacts}
\end{figure}

The variation of AoR with $V_{\max}$ at 10~vol\%, shown in Figure~\ref{fig:SimulationAndExperimentalPP2AOR_l5}, does not follow a monotonic trend. Instead, AoR increases up to $V_{\max}/\bar{V}_S = 1\%$ and decreases again for larger $V_{\max}$ values. This behavior can be explained by examining the number of wet contacts. Figure~\ref{fig:PP2_WetContacts} shows that the number of wet contacts reaches a maximum at $V_{\max}/\bar{V}_S = 1\%$ and decreases for both smaller and larger values. As a result, this case exhibits the strongest cohesive network and hence the highest AoR.

\begin{figure}[H]
	\centering
	\includegraphics[width=0.6\textwidth]{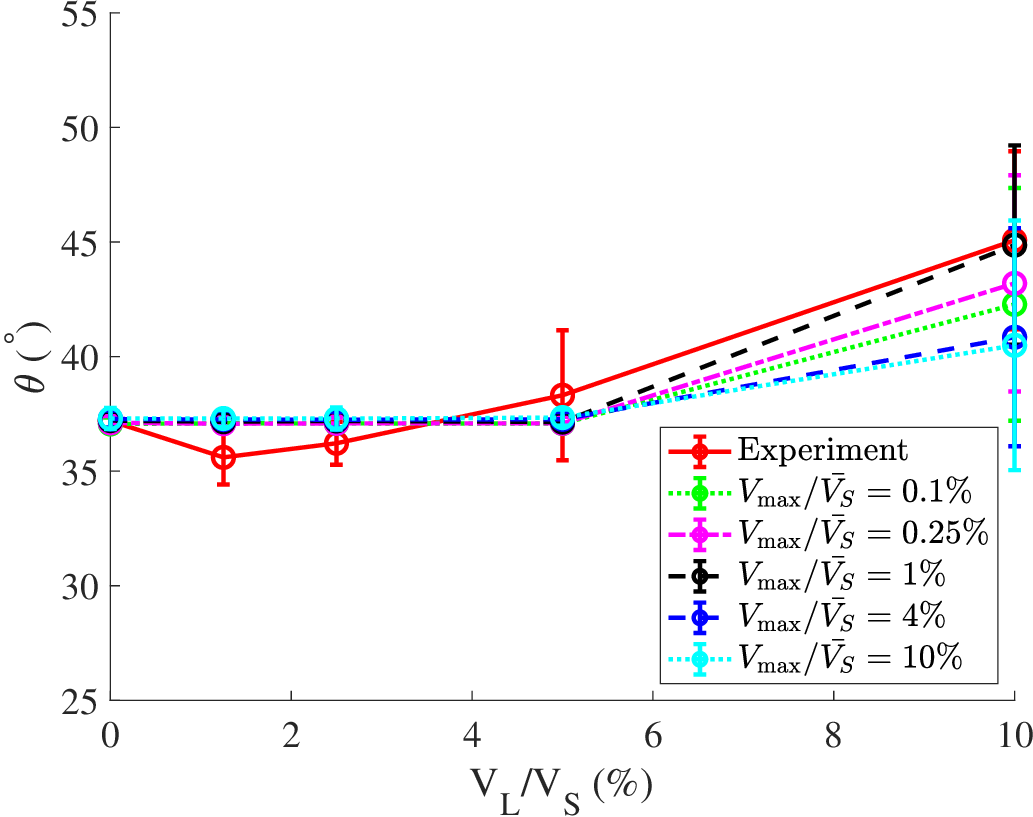}
	\caption{Experimental and simulated dynamic angle of repose (\(\theta\)) for PP2 at scaling factor \(l = 3\), comparing different values of maximum liquid bridge volume to determine the best match with experiments.}
	\label{fig:SimulationAndExperimentalPP2AOR_l3}
\end{figure}

To assess the influence of particle scaling, the same calibration procedure was repeated with a smaller particle scaling factor (\(l = 3\)) instead of 5. As seen in Figure~\ref{fig:SimulationAndExperimentalPP2AOR_l3}, the AoR values obtained at both scales are consistent, indicating that the calibrated parameters are robust and largely independent of particle scaling. The close agreement between the two scales confirms that the chosen parameter set reliably reproduces the macroscopic flow behavior.

\begin{figure}[H]
	\centering
	\renewcommand{\arraystretch}{1.2}
	\begin{tabular}{c@{\hspace{1.5pt}}c@{\hspace{1.5pt}}c@{\hspace{1.5pt}}c@{\hspace{1.5pt}}c@{\hspace{1.5pt}}c}
		\raisebox{0.5\height}{\small Scale 5} &
		\includegraphics[width=0.178\textwidth]{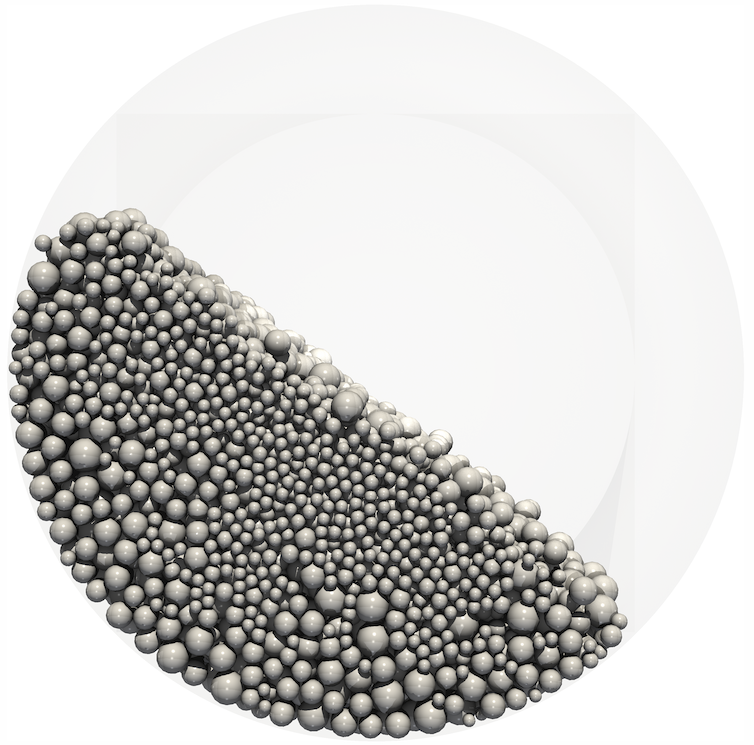} &
		\includegraphics[width=0.178\textwidth]{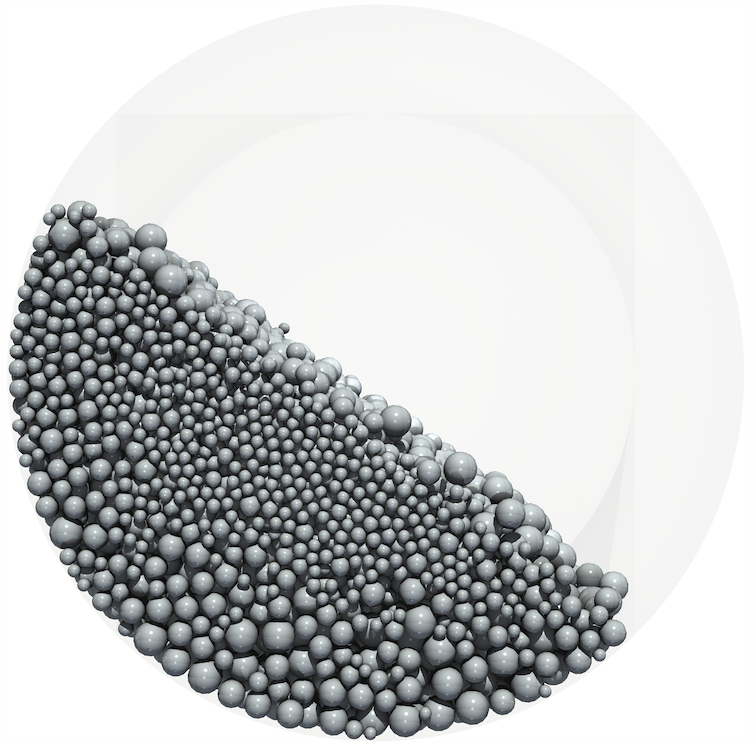} &
		\includegraphics[width=0.178\textwidth]{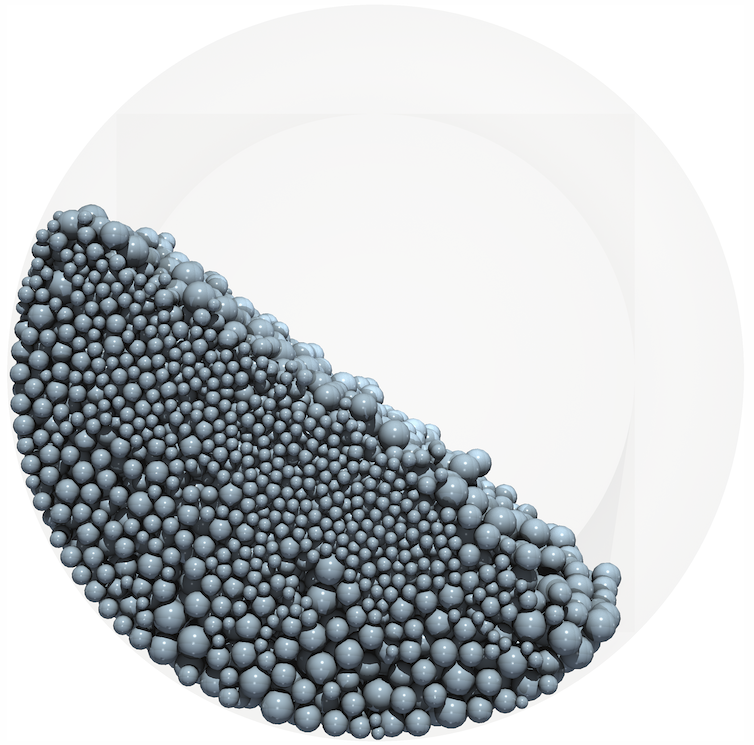} &
		\includegraphics[width=0.178\textwidth]{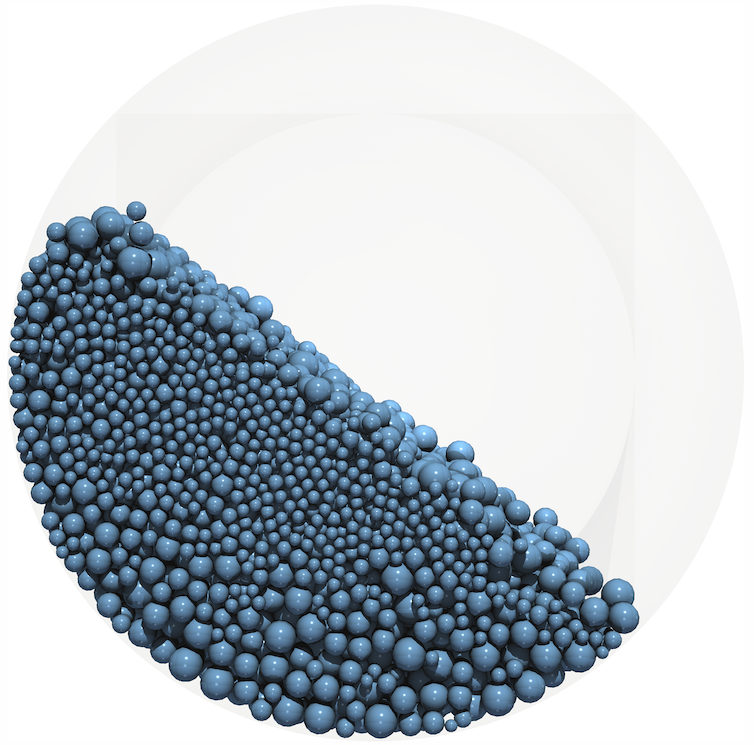} &
		\includegraphics[width=0.178\textwidth]{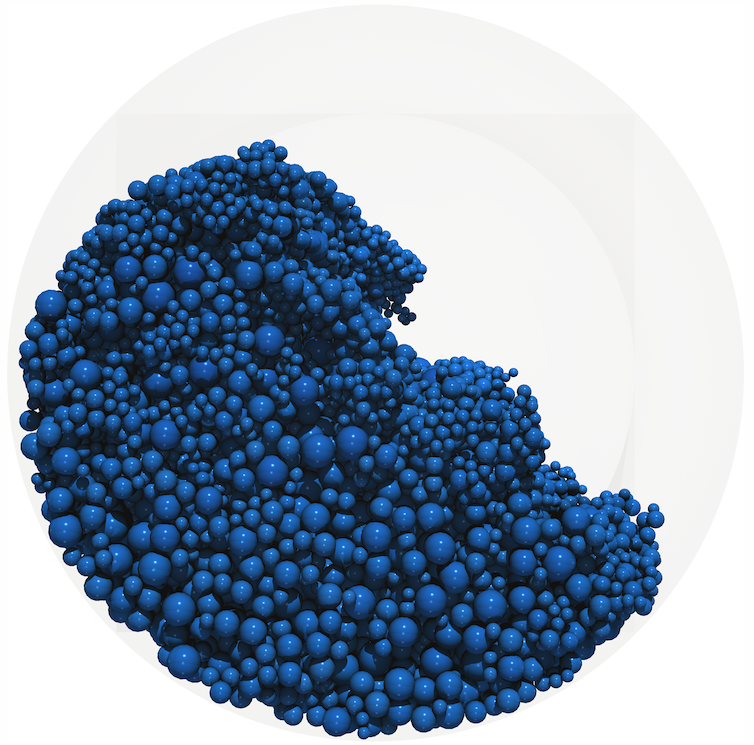} \\
		
		\raisebox{0.5\height}{\small Scale 3} &
		\includegraphics[width=0.178\textwidth]{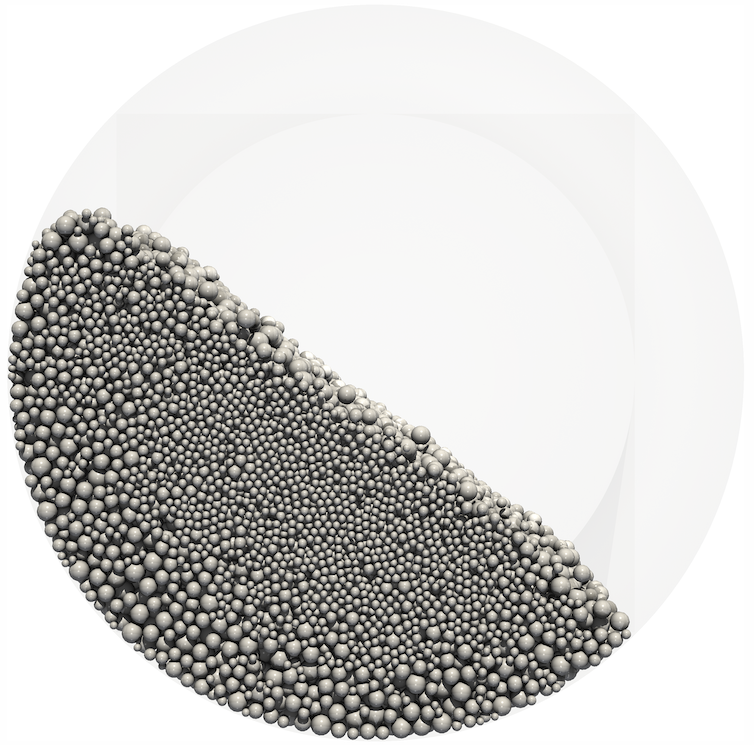} &
		\includegraphics[width=0.178\textwidth]{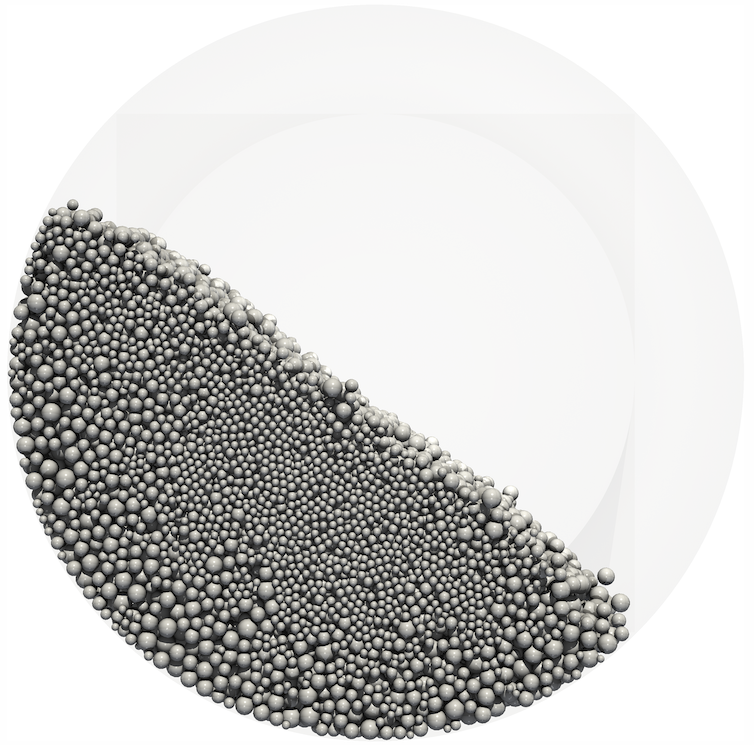} &
		\includegraphics[width=0.178\textwidth]{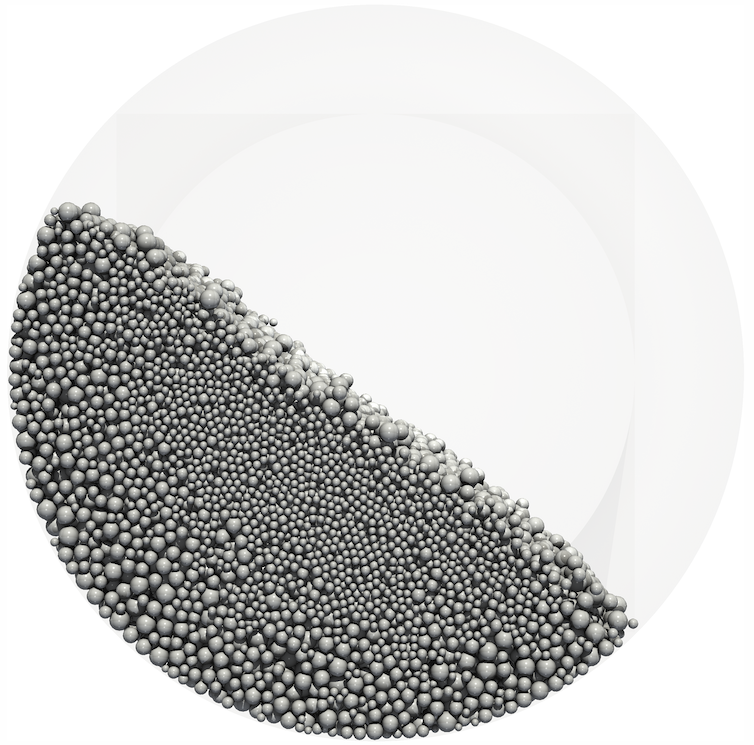} &
		\includegraphics[width=0.178\textwidth]{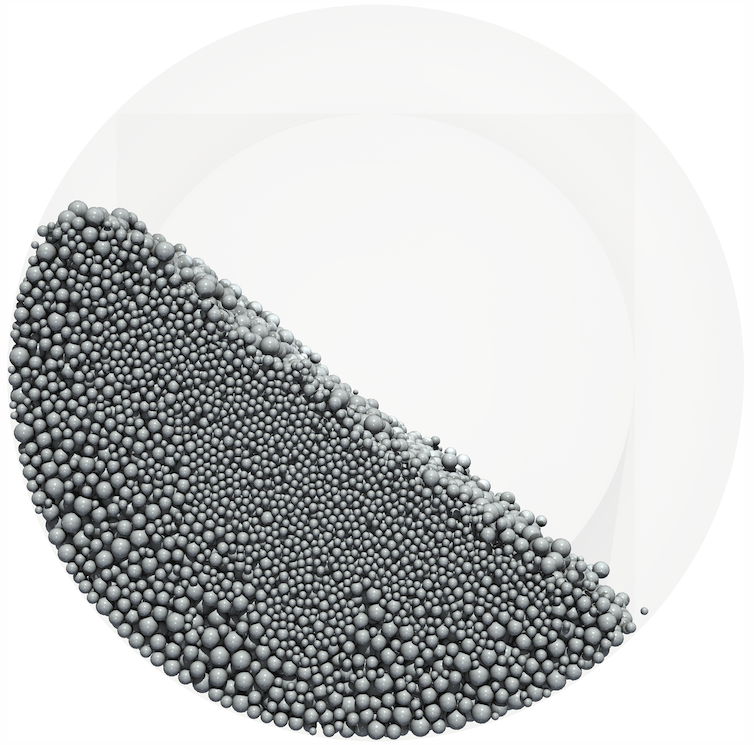} &
		\includegraphics[width=0.178\textwidth]{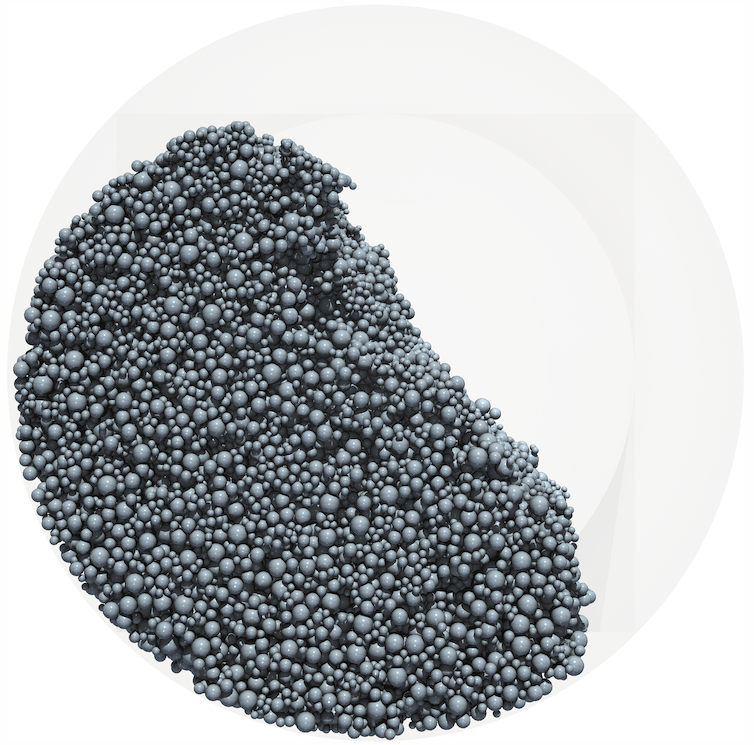} \\
		
		& \small 0 vol\% &
		\small 1.25 vol\% &
		\small 2.5 vol\% &
		\small 5 vol\% &
		\small 10 vol\% \\
	\end{tabular}
	\caption{Simulation snapshots of PP2 at two scaling factors and different liquid contents, showing the flow behavior. Coloring indicates the full liquid volume per particle (liquid film volume plus half the liquid bridge volume of the neighbouring contacts), ranging from $0$ (white) to \(15~\mu\text{L}\) (dark blue).}
	\label{fig:Scale3_Scale5_snapshots}
\end{figure}

While both scaling factors provided quantitative agreement with experimental AoR trends, qualitative differences appeared at higher liquid contents. As shown in Figure~\ref{fig:Scale3_Scale5_snapshots}, the flow behavior at low liquid contents (0--5~vol\%) is similar for both scales and matches experimental observations. However, at 10~vol\%, the flow at scale~3 exhibits smoother, more continuous motion, whereas the scale~5 simulation shows excessive clumping and reduced mobility. 

This behavior results from agglomerate formation at higher liquid contents. According to Reference~\cite{coetzee2019particle}, maintaining dynamic similarity requires the drum diameter to be at least 30 times larger than the characteristic particle size. Although this condition was met for individual scaled particles, the development of cohesive agglomerates effectively increased the representative particle size, thereby violating the similarity criterion at scale~5. Consequently, the smaller scale~3 simulation better preserved dynamic similarity under cohesive conditions.

In addition to these scaling effects, the flow structure also reveals a mild tendency toward particle-size segregation in the dry and weakly wetted cases, where larger particles accumulate more frequently near the free surface. This tendency diminishes as liquid content increases, consistent with cohesion suppressing percolation-driven size sorting. 

\begin{figure}[H]
	\centering
	\includegraphics[width=0.6\textwidth]{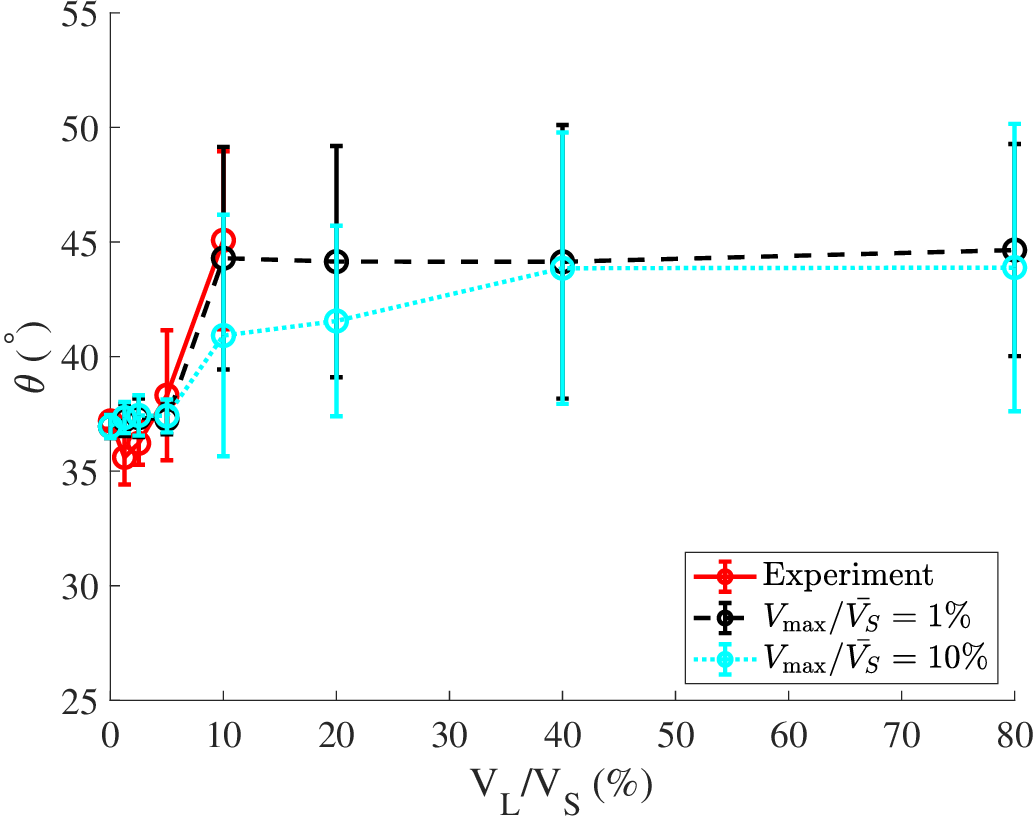}
	\caption{Experimental and simulated dynamic angle of repose (\(\theta\)) for PP2 ($l$ = 5) as a function of liquid content, comparing two values of maximum liquid bridge volume.}
	\label{fig:SimulationAndExperimentalAoRForHigherLiquid}
\end{figure}

Since the saturation point of cohesive interactions is not in the experimental range, simulations were extended to higher liquid contents (Figure~\ref{fig:SimulationAndExperimentalAoRForHigherLiquid}). The results confirm that \(V_{\min}\) governs the onset of liquid bridge formation and cohesion, while \(V_{\max}\) determines the saturation point. For \(V_{\max}/\bar{V}_S = 1\%\), saturation occurred near 10~vol\%, whereas for \(V_{\max}/\bar{V}_S = 10\%\), saturation was delayed until about 40~vol\%, consistent with the conceptual framework illustrated in Figure~\ref{fig:Schematic_AoR}.

\begin{figure}[H]
	\centering
	\renewcommand{\arraystretch}{1.2}
	\begin{tabular}{c@{\hspace{1.2pt}}c@{\hspace{1.2pt}}c@{\hspace{1.2pt}}c@{\hspace{1.2pt}}c@{\hspace{1.2pt}}c}
		\raisebox{0.5\height}{\small $\frac{V_{\max}}{\bar{V}_S}= 1\%$} &
		\includegraphics[width=0.17\textwidth]{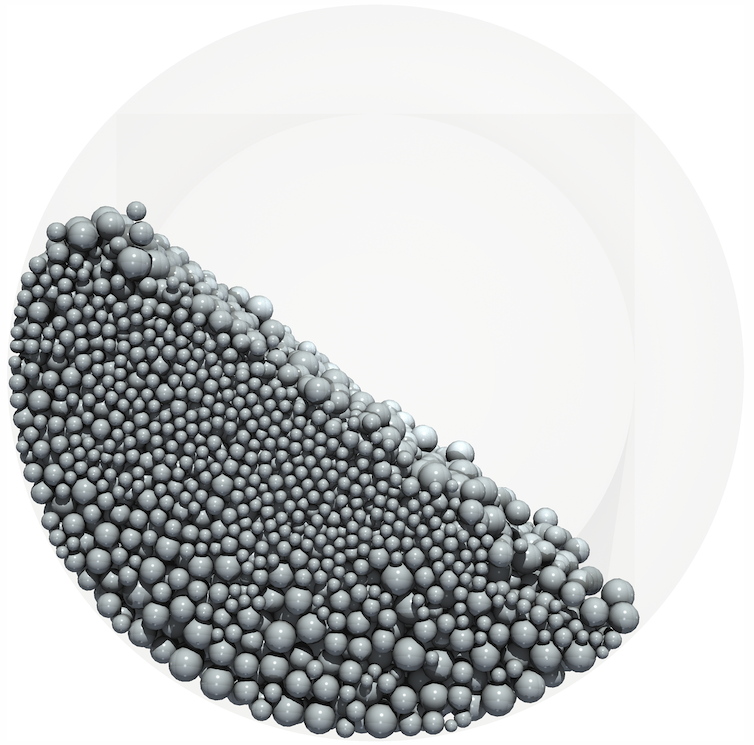} &
		\includegraphics[width=0.17\textwidth]{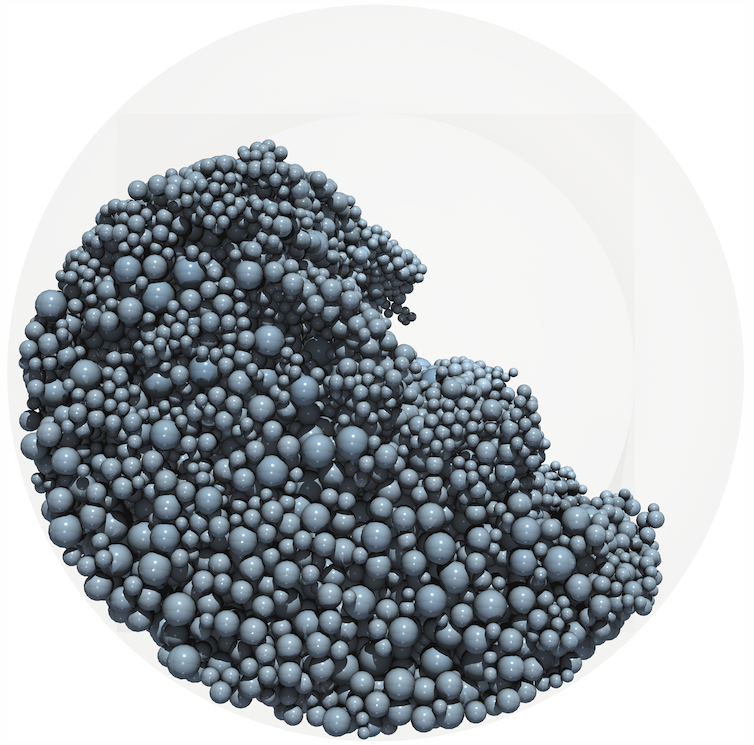} &
		\includegraphics[width=0.17\textwidth]{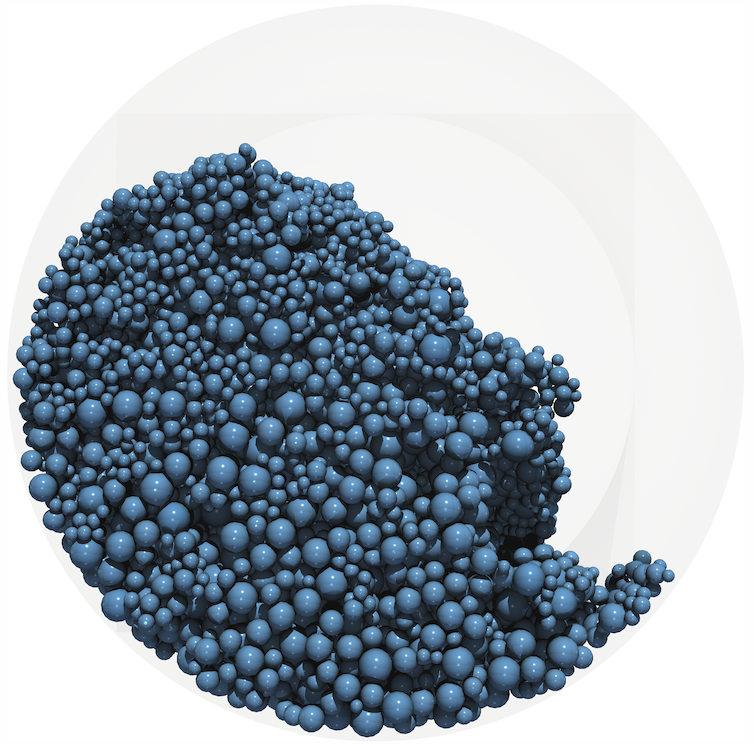} &
		\includegraphics[width=0.17\textwidth]{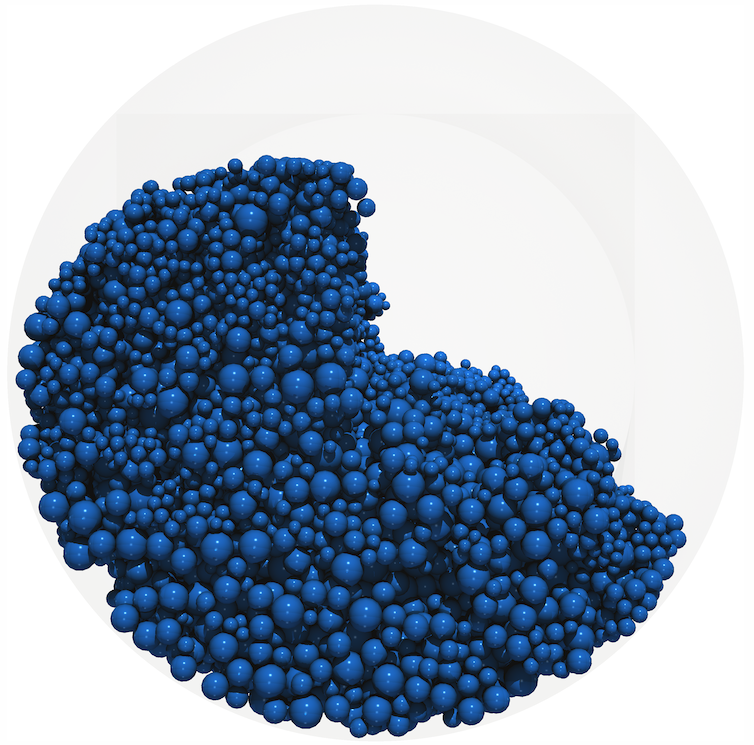} &
		\includegraphics[width=0.17\textwidth]{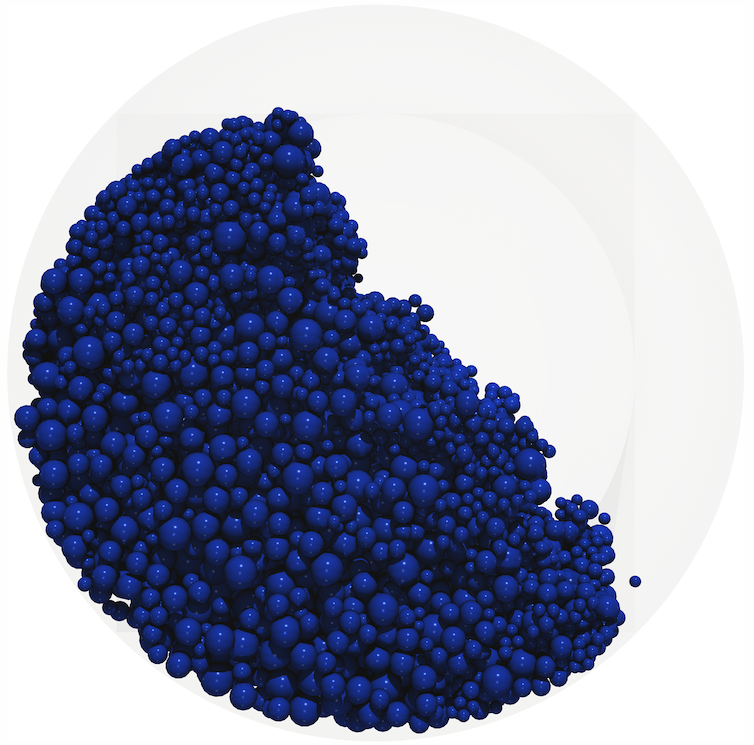} \\
		
		\raisebox{0.5\height}{\small $\frac{V_{\max}}{\bar{V}_S}= 10\%$} &
		\includegraphics[width=0.17\textwidth]{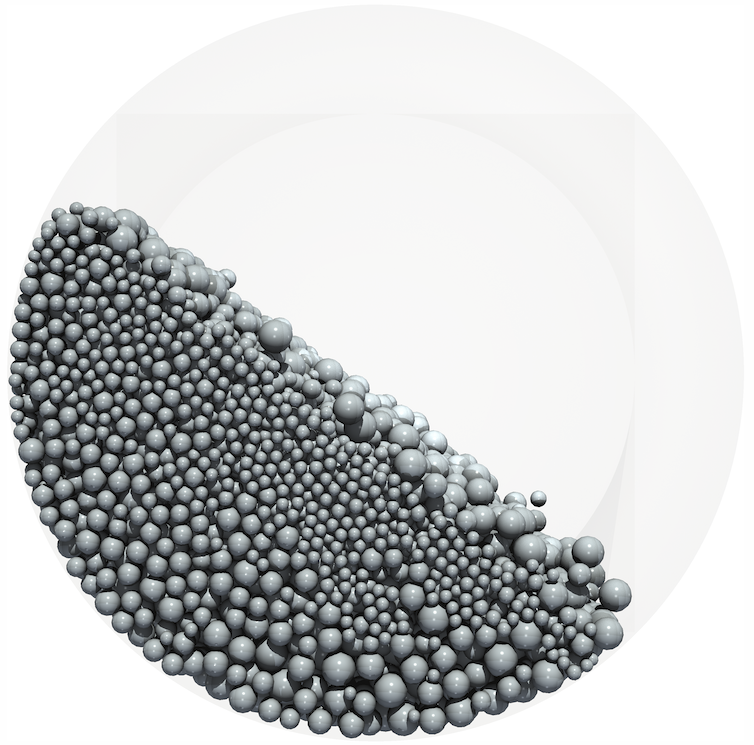} &
		\includegraphics[width=0.17\textwidth]{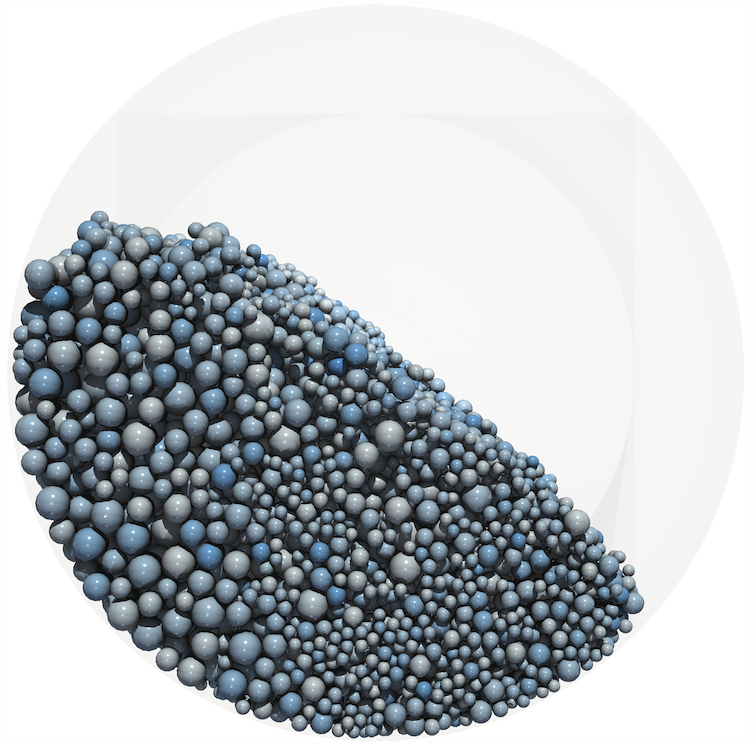} &
		\includegraphics[width=0.17\textwidth]{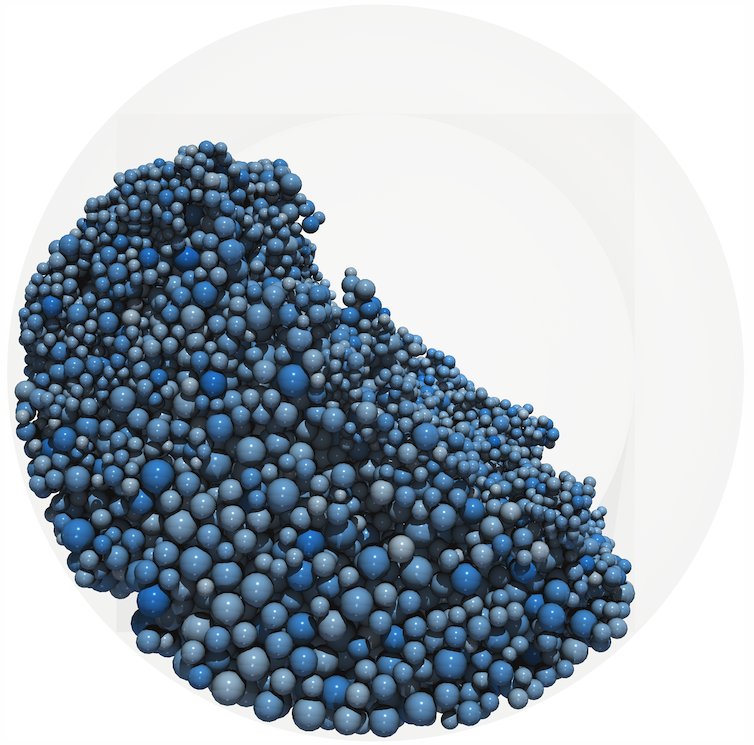} &
		\includegraphics[width=0.17\textwidth]{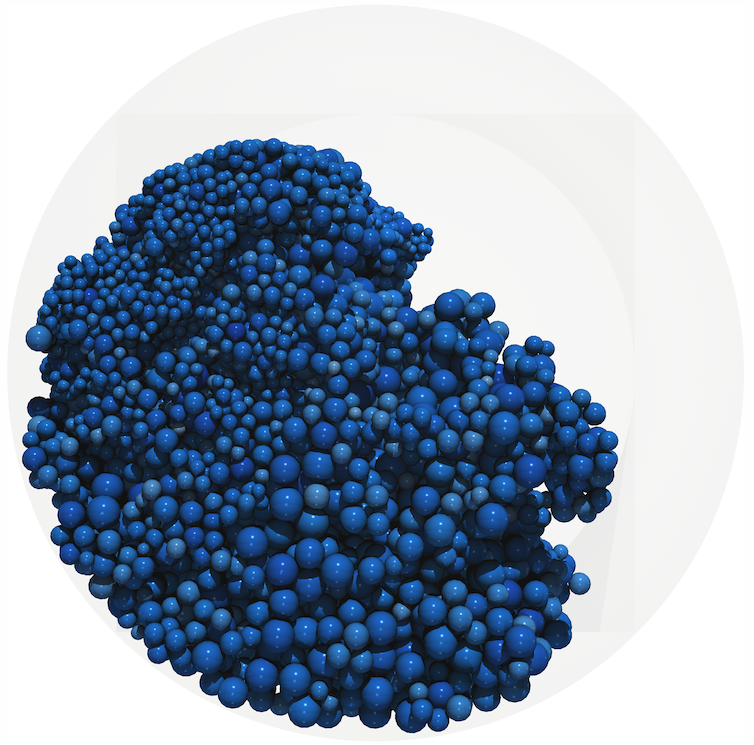} &
		\includegraphics[width=0.17\textwidth]{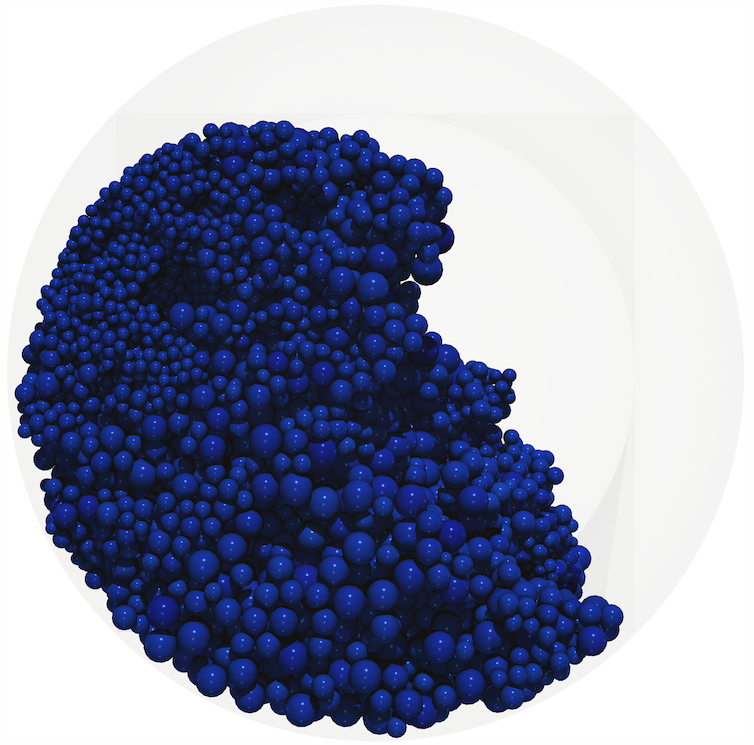} \\
		
		& \small 5 vol\% &
		\small 10 vol\% &
		\small 20 vol\% &
		\small 40 vol\% &
		\small 80 vol\% \\
	\end{tabular}
	\caption{Simulation snapshots of PP2 (\(l = 5\)) at two values of maximum liquid bridge volume and different liquid contents, showing the flow behavior. Coloring indicates the full liquid volume per particle, ranging from $0$ (white) to \(55~\mu\text{L}\) (dark blue).}
	\label{fig:Snapshots_TwoVmax}
\end{figure}

The simulation snapshots in Figure~\ref{fig:Snapshots_TwoVmax} visually confirm these trends. At low liquid contents, both cases behave similarly, but as liquid content increases, the larger bridge capacity (\(V_{\max} = 10\%\)) leads to stronger cohesion, more stable agglomerates, and reduced flowability. These results clearly demonstrate that \(V_{\min}\) controls the onset of cohesion, while \(V_{\max}\) governs the progression and saturation of cohesive interactions.

In summary, the calibrated parameters \(V_{\min}/\bar{V}_S = 10\%\) and \(V_{\max}/\bar{V}_S = 1\%\) successfully reproduced the experimental flow behavior of PP2. The analysis also highlighted that although coarser scaling (\(l = 5\)) can be used for efficient calibration, finer scaling (\(l = 3\)) is preferable when accurate representation of cohesive flow and agglomeration is required.

\subsection{Calibration of PP1}

\begin{figure}[H]
	\centering
	\renewcommand{\arraystretch}{1.2}
	\begin{tabular}{c@{\hspace{8pt}}c} % 2 columns
		\includegraphics[width=0.45\textwidth]{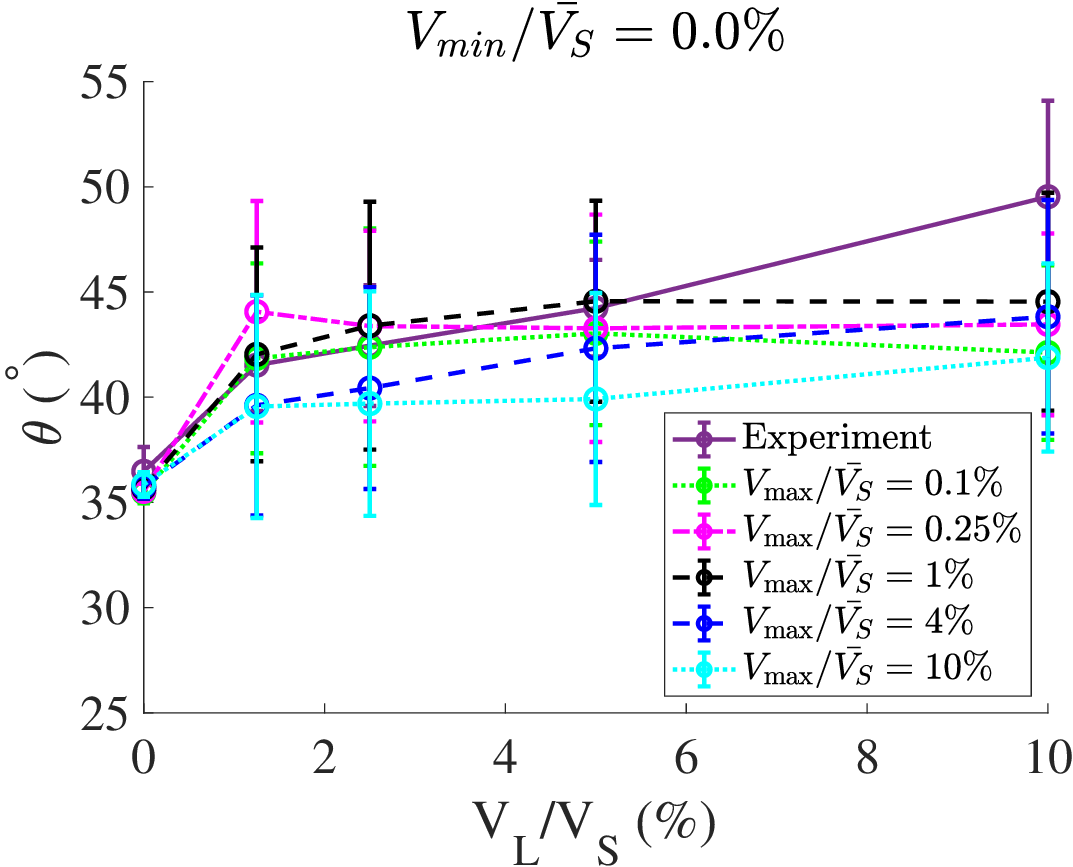} &
		\includegraphics[width=0.45\textwidth]{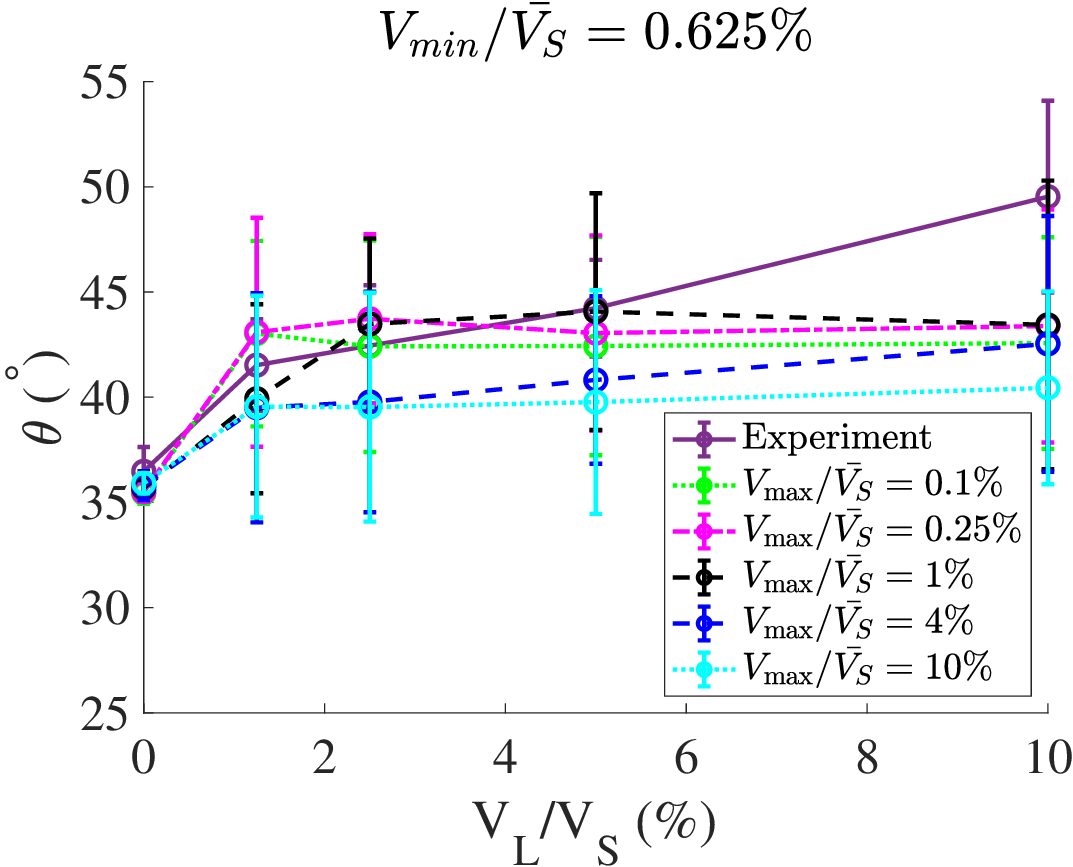} \\
		\includegraphics[width=0.45\textwidth]{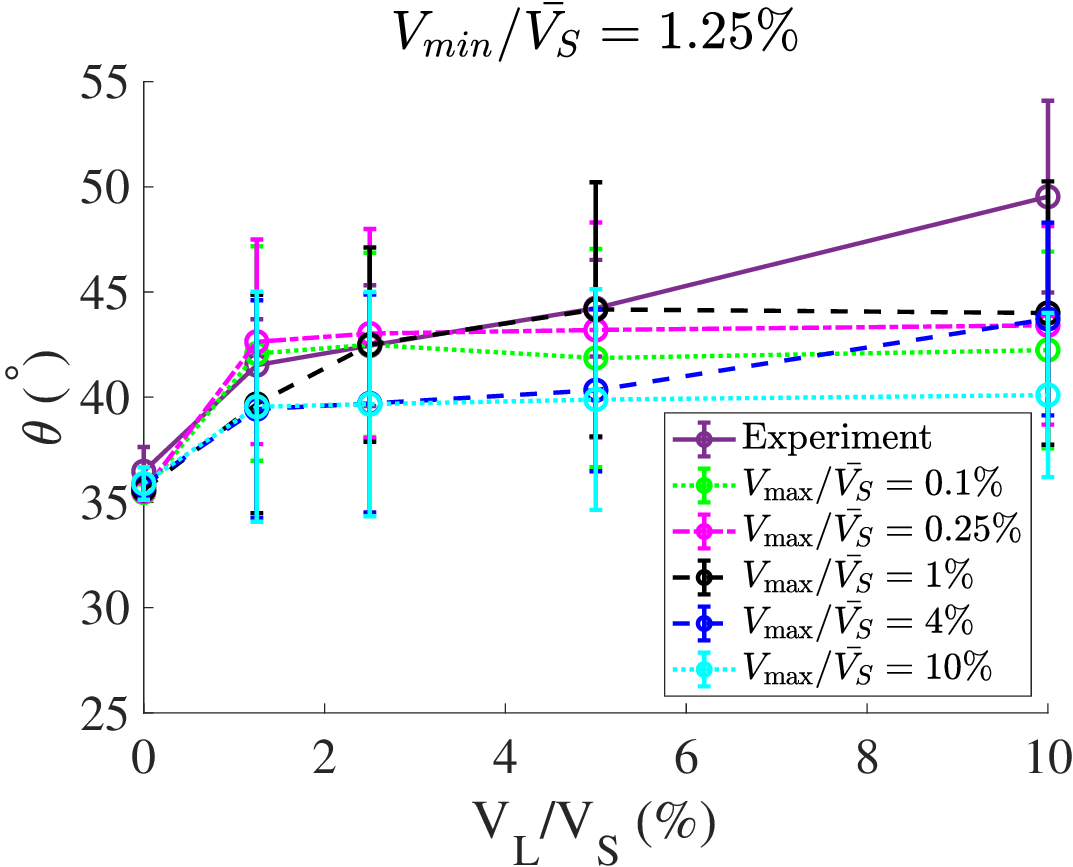} &
		\includegraphics[width=0.45\textwidth]{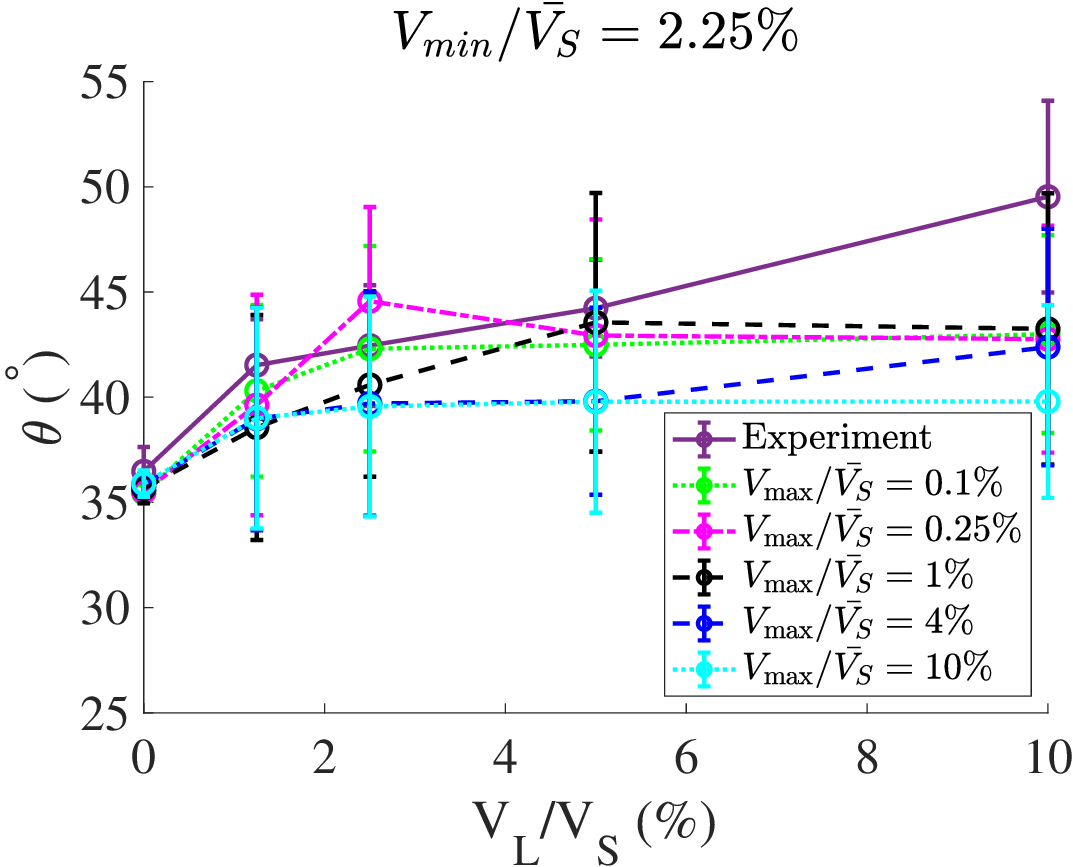} \\
	\end{tabular}
	\caption{Experimental and simulated dynamic angle of repose (\(\theta\)) for PP1 ($l$ = 5) as a function of liquid content to determine the best match with experiments. Each subfigure corresponds to a different minimum liquid film volume.}
	\label{fig:PP1_AOR_Vmin}
\end{figure}

Following the same calibration procedure applied to PP2, simulations were conducted for PP1 using the $V_{\min}$ and $V_{\max}$ values defined previously (Table~\ref{tab:PP1CalibrationParameters}). Figure~\ref{fig:PP1_AOR_Vmin} compares the simulated AoR values with the experimental results for four different values of $V_{\min}$. The best agreement with the experimental AoR data was achieved using $V_{\min}/\bar{V}_S = 0\%$ and $V_{\max}/\bar{V}_S = 1\%$, which provided a good match for liquid contents of 0, 1.25, 2.5, and 5\,vol\%. However, for the highest liquid content tested (10\,vol\%), no combination of $V_{\min}$ and $V_{\max}$ was able to reproduce the experimental AoR accurately.

This discrepancy is attributed to the same upscaling limitations previously observed with PP2. As the liquid content increases, agglomerates form and grow larger, effectively increasing the size of the cohesive structures compared to the dry particles. Since the simulation used a particle scaling factor \(l = 5\) to reduce computational cost, the ratio of drum diameter to the effective agglomerate diameter at high liquid content fell below the threshold (approximately 30) recommended for correctly reproducing granular flow dynamics \cite{coetzee2019particle,pourandi2024mathematical}. This led to a divergence in flow behavior between the simulation and the experiment at higher moisture levels. Reducing the scaling factor further would increase computational cost prohibitively, making it impractical for the current study.

\begin{figure}[H]
	\centering
	\renewcommand{\arraystretch}{1.2}
	\begin{tabular}{c@{\hspace{2pt}}c@{\hspace{2pt}}c@{\hspace{2pt}}c@{\hspace{2pt}}c@{\hspace{2pt}}c}
		\raisebox{0.5\height}{\small XY} &
		\includegraphics[width=0.18\textwidth]{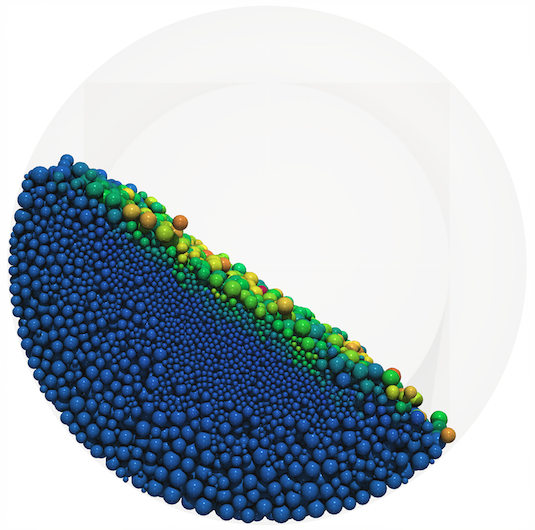} &
		\includegraphics[width=0.18\textwidth]{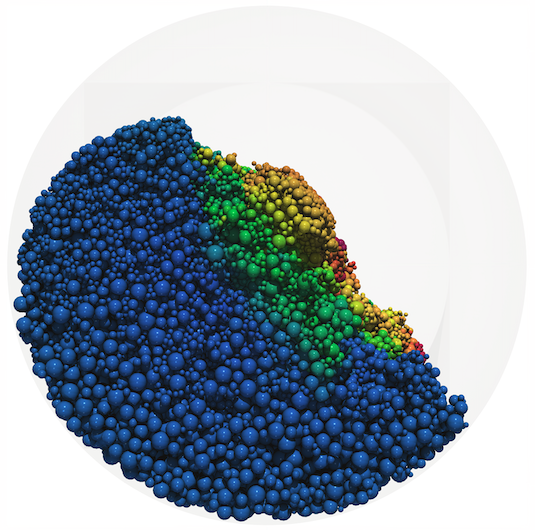} &
		\includegraphics[width=0.18\textwidth]{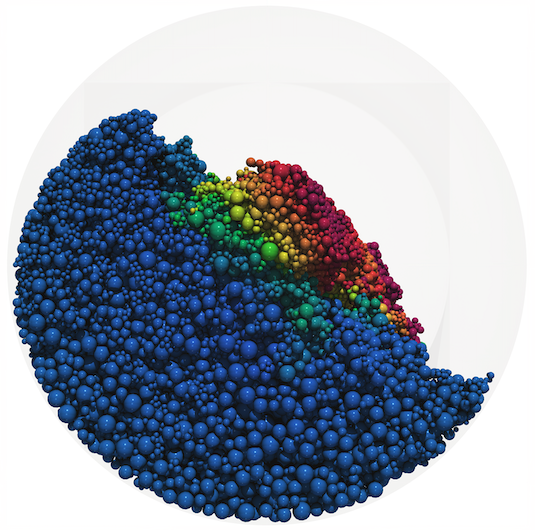} &
		\includegraphics[width=0.18\textwidth]{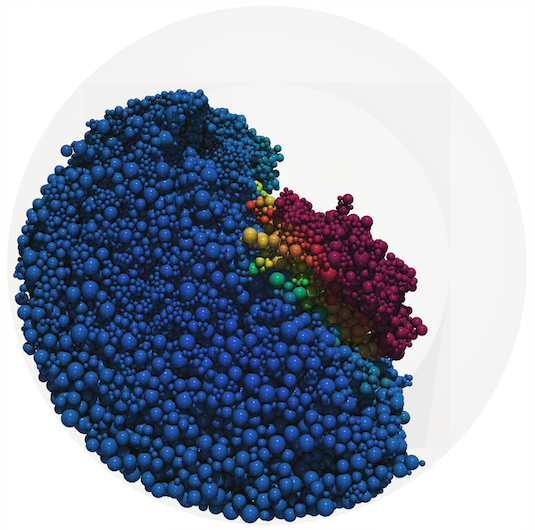} &
		\includegraphics[width=0.18\textwidth]{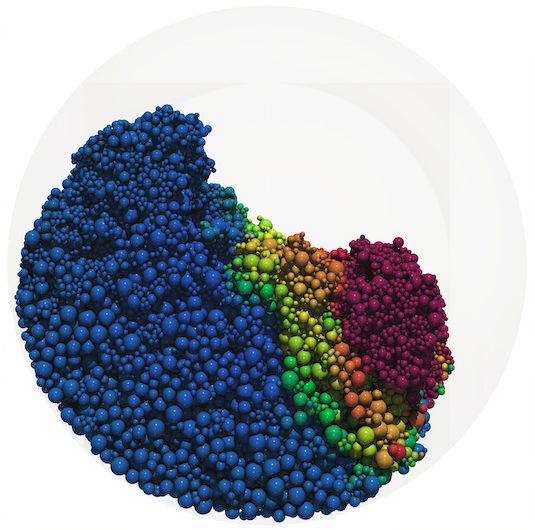} \\
		
		\raisebox{0.5\height}{\small YZ} &
		\includegraphics[width=0.18\textwidth]{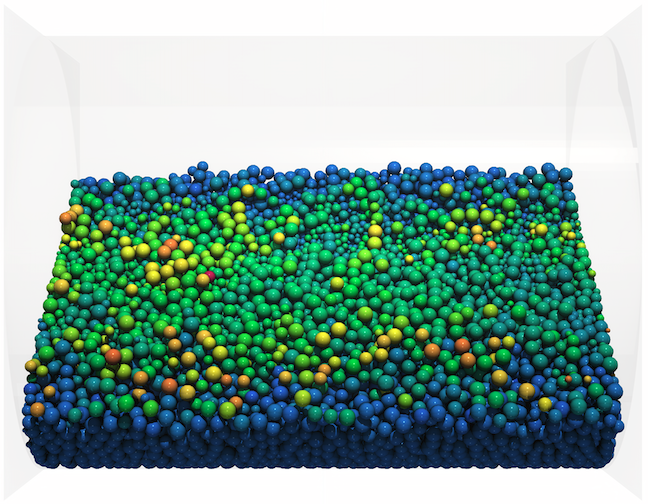} &
		\includegraphics[width=0.18\textwidth]{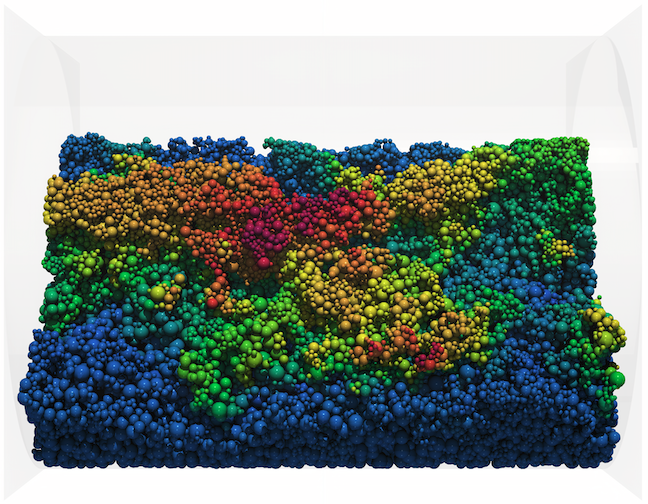} &
		\includegraphics[width=0.18\textwidth]{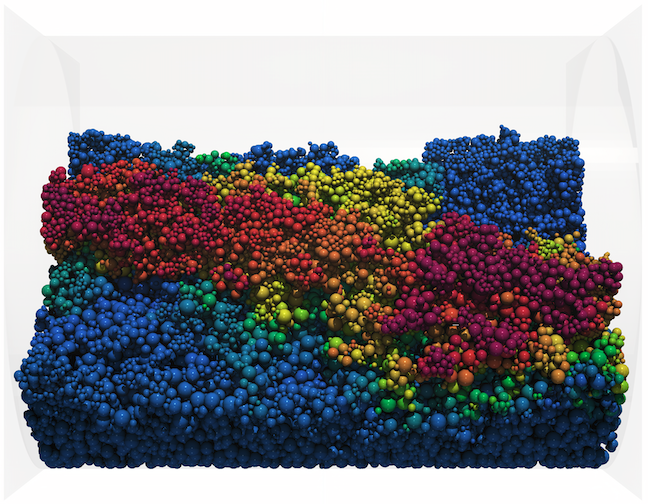} &
		\includegraphics[width=0.18\textwidth]{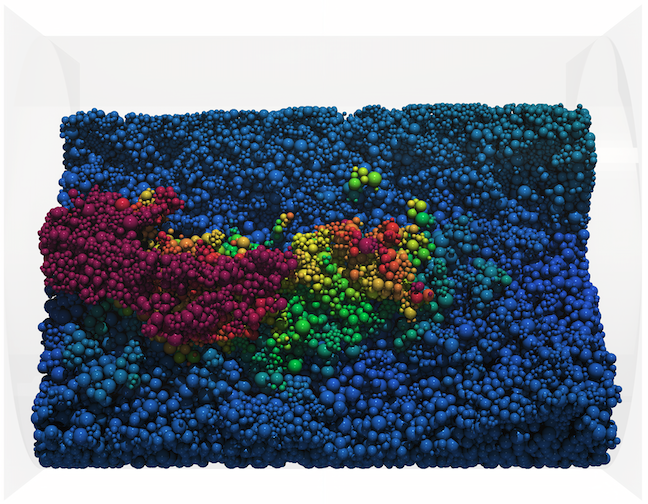} &
		\includegraphics[width=0.18\textwidth]{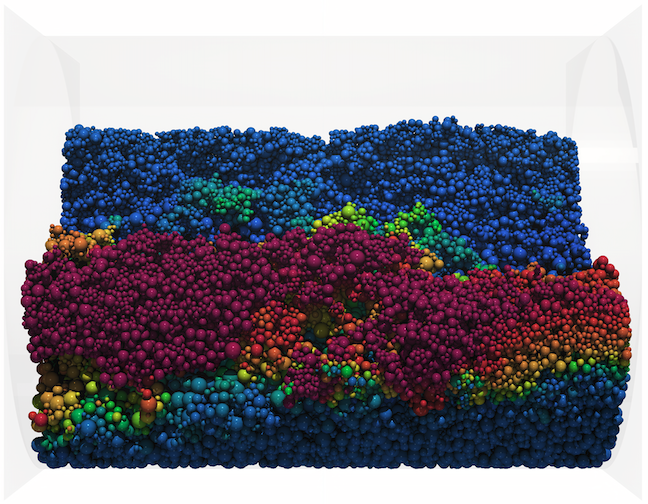} \\
		
		& \small 0 vol\% &
		\small 1.25 vol\% &
		\small 2.5 vol\% &
		\small 5 vol\% &
		\small 10 vol\% \\
	\end{tabular}
	\vspace{8pt}
	\includegraphics[height=1cm]{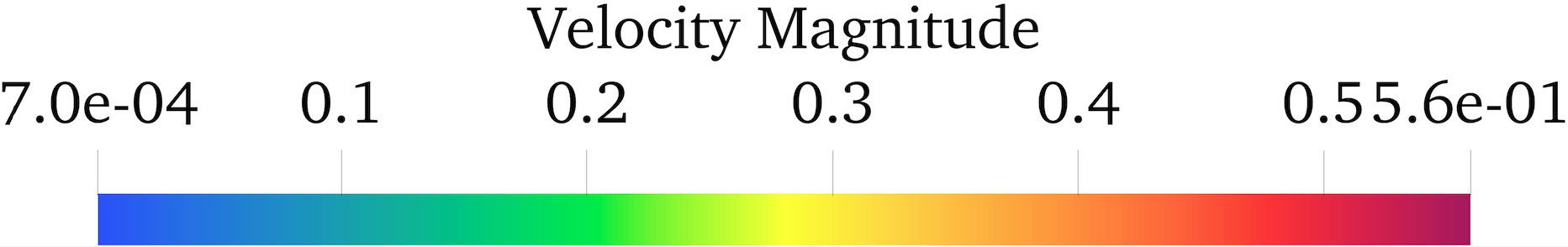}
	\caption{Simulation snapshots of PP1 (\(l = 5\)) at different liquid contents from the XY and YZ perspectives, showing the flow behavior.}
	\label{fig:PP1_velocity_snapshots}
\end{figure}

To further understand the differences between the simulated and experimental results, flow behavior was analyzed through simulation snapshots of PP1 at different liquid contents, shown in Figure~\ref{fig:PP1_velocity_snapshots}. Views from both the XY and YZ planes are presented to capture the full flow structure and particle arrangement.

At low liquid contents (0–5\,vol\%), the simulated flow remained consistent with experimental observations, showing stable flowing heaps and increasing AoR as liquid bridges formed and cohesion developed. However, at 10\,vol\% liquid content, the simulated particles exhibited pronounced agglomeration and localized flow disruption, which differed from the experimental flow patterns. This again reflects the scaling effects discussed earlier: the formation of large agglomerates in the simulation altered the flow regime due to an insufficient drum-to-agglomerate size ratio, limiting the model’s ability to capture the experimental behavior accurately at high moisture contents.

Consistent with PP2, visual segregation in PP1 is most apparent under dry/low-moisture conditions and becomes strongly suppressed at higher liquid contents as cohesive agglomerates form. We note this qualitatively here and leave a dedicated segregation study to future work.

\section{Conclusions and outlook}\label{sec:conclusions}
This study presented a calibration framework for Discrete Element Method (DEM) simulations of two wet industrial polypropylene powders, referred to as PP1 and PP2, operating in the pendular regime. The dynamic angle of repose (AoR) was used as the key bulk property for calibration due to its sensitivity to flow behavior under continuous motion, making it more suitable than static AoR for capturing the rheology of wet powders in stirred systems.

The calibration focused on adjusting two parameters of the liquid migration model—\(V_{\min}\) and \(V_{\max}\)—while surface tension and contact angle were taken from literature. A schematic framework was introduced to interpret how increasing liquid content transitions the system from partially to fully wet contacts, and thus to higher cohesivity. This conceptual approach supported the explanation of capillary cohesion effects on AoR trends.

For PP2, the simulation successfully matched the experimental AoR across all tested liquid contents. The best agreement was achieved using \(V_{\min}/\bar{V}_S = 10\%\) and \(V_{\max}/\bar{V}_S = 1\%\), capturing both the onset and saturation of cohesive effects. These parameters were robust across different scaling factors and remained consistent at higher moisture contents. However, although the AoR was well reproduced, mismatches in flow behavior were observed at higher liquid content when using larger particle scaling factors. These discrepancies were reduced when the scaling factor was lowered from 5 to 3, improving the agreement between simulated and experimental flow patterns.

For PP1, the same calibration approach reproduced the experimental AoR up to 5\,vol\% liquid content. At 10\,vol\%, however, the simulated AoR deviated from the experimental values due to the formation of large agglomerates. These agglomerates effectively reduced the drum-to-particle size ratio in the simulation, altering the cohesive flow regime and limiting calibration accuracy at higher liquid contents.

Overall, the study confirms that dynamic AoR is a reliable and practical calibration target for simulating the behavior of wet industrial powders using DEM. The physically interpretable parameters \(V_{\min}\) and \(V_{\max}\) offer a practical route for tuning DEM models of lightly wetted materials and for linking micro-scale liquid redistribution to bulk rheology. The findings support AoR-based tuning of liquid-bridge parameters and highlight both its strengths and limitations when scaling is used to manage computational cost.

Future work should explore strategies to reduce scaling artefacts in cohesive regimes, for example, by using smaller experimental drums that allow simulations with little to no upscaling, thereby preserving realistic drum-to-agglomerate size ratios. A complementary direction is the development of agglomeration-aware scaling laws, in which the particle scaling reflects not only the size of individual particles but also the expected size of cohesive clusters that form in wet systems. Such approaches may enable more realistic and computationally efficient DEM simulations of strongly cohesive powders.

Finally, extending the framework beyond the pendular regime—by incorporating funicular and capillary wetting states—and combining AoR-based calibration with additional bulk tests could broaden its applicability to a wider range of industrial wet-granular processes.

\section*{Acknowledgment}

This research is part of the Industrial Dense Granular Flows project that received funding from the Dutch Research Council (NWO) in the framework of the ENW PPP Fund for the top sectors and from the Ministry of Economic Affairs in the framework of the “PPS-Toeslagregeling”.

Authors sincerely thank Professor Stefan Luding for valuable discussions on the topic.

\bibliography{ref}

\end{document}